\documentclass[twocolumn]{aastex6}

\usepackage{hyperref}
\usepackage{graphicx}
\usepackage{natbib}
\usepackage{amsmath,amsthm,amssymb}
\usepackage{url}
\usepackage{morefloats}

\def\WISE{\textit{WISE}}
\def\WD{\textit{WISE}/DEIMOS}

\protected\def\Lsed{\ifmmode \,\mathcal{L}_{\mathrm{SED}}\else $\mathcal{L}_{\mathrm{SED}}$\fi}

\protected\def\picometer{\ifmmode \,\operatorname{pm}\else $\operatorname{pm}$\fi}
\protected\def\nm{\ifmmode \,\operatorname{nm}\else $\operatorname{nm}$\fi}
\protected\def\micron{\ifmmode \,\operatorname{\mu m}\else $\operatorname{\mu m}$\fi}
\protected\def\mm{\ifmmode \,\operatorname{mm}\else $\operatorname{mm}$\fi}
\protected\def\meter{\ifmmode \,\operatorname{m}\else $\operatorname{m}$\fi}
\protected\def\km{\ifmmode \,\operatorname{km}\else $\operatorname{km}$\fi}
\protected\def\au{\ifmmode \,\operatorname{AU}\else $\operatorname{AU}$\fi}
\protected\def\pc{\ifmmode \,\operatorname{pc}\else $\operatorname{pc}$\fi}
\protected\def\kpc{\ifmmode \,\operatorname{kpc}\else $\operatorname{kpc}$\fi}
\protected\def\Mpc{\ifmmode \,\operatorname{Mpc}\else $\operatorname{Mpc}$\fi}
\protected\def\rsun{\ifmmode \,\operatorname{R_\odot}\else $\operatorname{R_\odot}$\fi}
\protected\def\Rsun{\ifmmode \,\operatorname{R_\odot}\else $\operatorname{R_\odot}$\fi}

\protected\def\second{\ifmmode \,\operatorname{sec}\else $\operatorname{sec}$\fi}
\protected\def\yr{\ifmmode \,\operatorname{yr}\else $\operatorname{yr}$\fi}
\protected\def\Gyr{\ifmmode \,\operatorname{Gyr}\else $\operatorname{Gyr}$\fi}

\protected\def\eV{\ifmmode \,\operatorname{eV}\else $\operatorname{eV}$\fi}
\protected\def\keV{\ifmmode \,\operatorname{keV}\else $\operatorname{keV}$\fi}
\protected\def\MeV{\ifmmode \,\operatorname{MeV}\else $\operatorname{MeV}$\fi}
\protected\def\GeV{\ifmmode \,\operatorname{GeV}\else $\operatorname{GeV}$\fi}
\protected\def\TeV{\ifmmode \,\operatorname{TeV}\else $\operatorname{TeV}$\fi}

\protected\def\Lsun{\ifmmode \,\operatorname{L_\odot}\else $\operatorname{L_\odot}$\fi}
\protected\def\lsun{\ifmmode \,\operatorname{L_\odot}\else $\operatorname{L_\odot}$\fi}
\protected\def\Watt{\ifmmode \,\operatorname{W}\else $\operatorname{W}$\fi}
\protected\def\nW{\ifmmode \,\operatorname{nW}\else $\operatorname{nW}$\fi}

\protected\def\kJy{\ifmmode \,\operatorname{kJy}\else $\operatorname{kJy}$\fi}
\protected\def\Jy{\ifmmode \,\operatorname{Jy}\else $\operatorname{Jy}$\fi}
\protected\def\mJy{\ifmmode \,\operatorname{mJy}\else $\operatorname{mJy}$\fi}
\protected\def\microJy{\ifmmode \,\operatorname{\mu Jy}\else $\operatorname{\mu Jy}$\fi}
\protected\def\nJy{\ifmmode \,\operatorname{nJy}\else $\operatorname{nJy}$\fi}

\protected\def\Mag{\ifmmode \,\operatorname{mag}\else $\operatorname{mag}$\fi}

\protected\def\deg{\ifmmode ^{\circ}\else $^{\circ}$\fi}
\protected\def\arcsec{\ifmmode ^{\prime\prime}\else $^{\prime\prime}$\fi}

\protected\def\arcsecT{\ifmmode \,\operatorname{arcsec}\else $\operatorname{arcsec}$\fi}
\protected\def\arcmin{\ifmmode ^{\prime}\else $^{\prime}$\fi}

\protected\def\arcminT{\ifmmode \,\operatorname{arcmin}\else $\operatorname{arcmin}$\fi}
\protected\def\sr{\ifmmode \,\operatorname{sr}\else $\operatorname{sr}$\fi}

\newcommand{\code}[1]{\texttt{#1}}

\protected\def\d{\ifmmode \operatorname{d}\else
    $\operatorname{d}$\fi}
\protected\def\e{\ifmmode \operatorname{e}\else
    $\operatorname{e}$\fi}

\def\apjl{{ApJ}}

\def\apjs{{ApJS}}

\def\apjref#1;#2;#3;#4 {\par\pp#1, {#2}, #3, #4 \par}

\shorttitle{WISE Galaxy Luminosity Function}
\shortauthors{Lake et al.}

\begin{document}

\title{The 2.4 $\mu$m Galaxy Luminosity Function as Measured Using \WISE. II. Sample Selection}
\author{S.~E.~Lake\altaffilmark{1}, E.~L.~Wright\altaffilmark{1}, R.~J.~Assef\altaffilmark{2}, T.~H.~Jarrett\altaffilmark{3}, S.~Petty\altaffilmark{4}, S.~A.~Stanford\altaffilmark{5,6}, D.~Stern\altaffilmark{7},  C.-W.~Tsai\altaffilmark{1,7}}

\altaffiltext{1}{Physics and Astronomy Department, University of California, Los Angeles, CA 90095-1547}
\altaffiltext{2}{N\'ucleo de Astronom\'ia de la Facultad de Ingenier\'ia y Ciencias, Universidad Diego Portales, Av. Ej\'ercito Libertador 441, Santiago, Chile}
\altaffiltext{3}{Astronomy Department University of Cape Town Private Bag X3 Rondebosch 7701 Republic of South Africa}
\altaffiltext{4}{NorthWest Research Associates
4118 148th Ave NE
Redmond, WA 98052-5164}
\altaffiltext{5}{Department of Physics, University of California, Davis, CA 95616}
\altaffiltext{6}{Institute of Geophysics and Planetary Physics, Lawrence Livermore National
Laboratory, Livermore CA 94551}
\altaffiltext{7}{Jet Propulsion Laboratory, California Institute of
Technology, 4800 Oak Grove Dr., Pasadena, CA 91109}

\email{lake@physics.ucla.edu}

\begin{abstract}
The \WISE\ satellite surveyed the entire sky multiple times in four infrared (IR) wavelengths \citep[3.4, 4.6, 12, and $22\micron$;][]{Wright:2010}. 
This all-sky IR photometric survey makes it possible to leverage many of the large publicly available spectroscopic redshift surveys to measure galaxy properties in the IR. 
While characterizing the cross-matching of \WISE\ data to a single survey is a straightforward process, doing it with six different redshift surveys takes a fair amount of space to characterize adequately, because each survey has unique caveats and characteristics that need addressing. 
This work describes a data set that results from matching five public redshift surveys with the AllWISE data release, along with a reanalysis of the data described in \cite{Lake:2012}. 
The combined data set has an additional flux limit of $80\microJy$ ($19.14\operatorname{AB\ mag}$) in \WISE's W1 filter imposed in order to limit it to targets with high completeness and reliable photometry in the AllWISE data set. 
Consistent analysis of all of the data is only possible if the color bias discussed in \cite{Ilbert:2004} is addressed \citep[for example: the techniques explored in][]{Lake:2017a}. 
The sample defined herein is used in this paper's companion paper, \cite{Lake:2018b}, to measure the luminosity function of galaxies at $2.4\micron$ rest frame wavelength, and the selection process of the sample is optimized for this purpose.
\end{abstract}

\keywords{catalogs, surveys, galaxies: statistics}


\section{Introduction}
The astronomy community has an embarrassment of riches when it comes to the depth and breadth of its publicly available catalog of data, both photometric (Sloan Digital Sky Survey [SDSS], Two Micron All Sky Survey [2MASS], Galaxy Evolution Explorer [GALEX], Widefield Infrared Survey Explorer [WISE], to name a few), and spectroscopic (for example, 6dF Galaxy Survey [6dFGS], SDSS, Galaxy and Mass Assembly [GAMA]). 
As tempting as it is to combine all of the available data into a unified measurement of the luminosity function (LF) of galaxies, there is no single standard for how targets are selected for measurement of spectroscopic redshifts or how the resulting data is characterized. 
So any effort to analyze a data set that is a synthesis of many data sets requires a careful consideration of whether the target selection processes are sufficiently similar to be modeled in a unified way, and characterization of the resulting data set after all quality cuts are made. 

We first encountered these difficulties when we made the decision to augment the data in our own small survey, \WD\ \citep[][]{Lake:2012}, with publicly available spectroscopic redshift surveys to measure the luminosity function of galaxies at a wavelength of $2.4\micron$. 
The plan was to analyze the data from multiple public surveys both separately and together to get a good grasp on systematics, increase sample size, and minimize cosmic variance.
To that end, we selected five additional surveys, with the intention that no one survey should be unique in any redshift range, that were as close to \WD\ in sample selection as possible.
Critically, the surveys had to be as close to flux limited in one digital imaging filter as possible, ruling out surveys like: SDSS Luminous Red Galaxy Sample \citep[color cut][]{Eisenstein:2001}, the 2dF Galaxy Redshift Survey \cite[2dFGRS; photographic magnitude][]{Colless:2003}, and the Deep Extragalactic Evolutionary Probe 2 survey \cite[DEEP2; color cut][]{Eisenstein:2001}. 

This paper contains a description of the sample selection process, and a characterization of the same, for spectroscopically measured redshift catalogs of galaxies pulled from six different surveys and crossmatched, wherever possible, to additional photometric information from SDSS data release 10 (SDSS-DR10), the 2-Micron All Sky Survey (2MASS), and the AllWISE Source and Reject Catalogs. 
The spectroscopic galaxy surveys tapped are: the 6dFGS Data Release 3 $K_s$ selected sample \citep{Jones:2009}, SDSS Main Galaxy Sample \citep{Strauss:2002, Ahn:2014}, GAMA data release 2 \citep{Liske:2015}, the AGN and Galaxy Evolution Survey (AGES) \citep{AGES}, the zCOSMOS 10k-Bright Spectroscopic Sample \citep{Lilly:2009}, and a reanalysis of the \WD\ survey. 
What all of these surveys have in common is that their target selection processes are driven, primarily, by observed flux in one channel: 2MASS $K_s$, SDSS $r$, SDSS $r$, Hubble F814W (approximately $I$), NOAO Deep Wide Field Survey (NDWFS) $I$, and \WISE\ W1, respectively. 
Roughly, the resulting data sets synthesized split up into three low-$z$ high $\Omega$ surveys (6dFGS, SDSS, and GAMA) and three high-$z$ low $\Omega$ surveys (AGES, zCOSMOS, and \WD).
This means that no one survey has a monopoly on the information from any redshift range, though the large number of targets in SDSS means that this information is only available if the surveys are analyzed separately before synthesizing them.
The simple selection process these surveys share leaves room, computationally, for the imposition of a further flux cut in W1 for the combined catalog without significantly increasing the complexity of the selection process.

A measurement of the LF using the data described herein takes place in a companion paper to this one, \citep[][from here on LW17III]{Lake:2018b}. 
The technique needed to account for the biases in such a diverse group of redshift surveys simultaneously is to broaden the concept of the LF to be a density over the galaxies' entire spectral energy distribution (SED), as described in \cite[][from here on LW17I]{Lake:2018a}. 
Two necessary components of that process are the mean and spectral covariance of galaxy spectra. 
This paper and its companion, LW17III, make use of the mean and covariance of galaxy SEDs as measured in \cite{Lake:2016}. 
The techniques developed allow us to address the SED dependent completeness concerns raised in \cite{Ilbert:2004}, and make the minimum cuts to the data needed in the process. 
In this work, the mean SED is used to cut galaxies that are likely low luminosity outliers in luminosity-redshift space, caused by contamination, as well as to compute curves bounding the regions in luminosity-redshift space where SED variety completeness is nearly constant.

In spite of our efforts to make the combined data set as homogeneous as possible, the character of each component survey is still distinct enough that they require separate consideration in any application.
With that in mind, we have structured this paper to reflect that fact, with each survey getting its own separate section for consideration, in spite of the repetition this causes.

The layout of this paper is as follows: Section~\ref{sec:sel} describes the data sets chosen and all cuts made to the sets. 
Each data set has its own subsection where details peculiar to it are described. 
The effects of the primary cuts on the data are demonstrated in graphs that are not completeness corrected in order to show what physical parameters, primarily redshift and luminosity, measurements based on the data set will be most sensitive to. 
Section~\ref{sec:dat} contains excerpts from the machine readable tables of the selected data published with this work. Finally, Section~\ref{sec:disc} contains concluding remarks.

The cosmology used in this paper is based on the WMAP 9 year $\Lambda$CDM cosmology \citep{Hinshaw:2013}\footnote{\url{http://lambda.gsfc.nasa.gov/product/map/dr5/params/lcdm_wmap9.cfm}}, with flatness imposed, yielding: $\Omega_M = 0.2793,\ \Omega_\Lambda = 1 - \Omega_M$, and $H_0 = 70 \km \second^{-1} \Mpc^{-1}$ (giving Hubble time $t_H = H_0^{-1} = 13.97\operatorname{Gyr}$, and  Hubble distance $D_H =  c t_H = 4.283 \operatorname{Gpc}$). 
All magnitudes will be in the AB magnitude system, unless otherwise specified. 
In cases where the source data was in Vega magnitudes and a conversion to the AB system was provided in the documentation, they were used (2MASS\footnote{\url{http://www.ipac.caltech.edu/2mass/releases/allsky/faq.html\#jansky}} and AllWISE\footnote{\url{http://wise2.ipac.caltech.edu/docs/release/allsky/expsup/sec4_4h.html\#WISEZMA}}). 
For the surveys without obviously documented Vega/AB magnitude offsets (NDWFS\footnote{\url{http://www.noao.edu/noao/noaodeep/}}, SDWFS\footnote{\url{http://irsa.ipac.caltech.edu/data/SPITZER/SDWFS/}}) we performed the conversion using those provided in \cite{AGES}. 

\section{Data Selection and Characterization} \label{sec:sel}
The defining data set of this paper is the \WISE\ W1 selected survey described in \cite{Lake:2012}, hereafter \WD. 
The biggest advantage of \WD\ is that the target selection function is extremely simple, driven at the faint end entirely by the target's flux at $3.4\micron$, W1 filter. 
The disadvantage is that the sample size is relatively small (\mbox{$N \sim 200$}). 
The smallness of the \WD\ sample, and AllWISE's sky coverage, is what drove the decision to leverage the existing catalog of redshift surveys. 
We imposed a W1 flux limit of $80\microJy\ (19.14\operatorname{AB\ mag})$ on all surveys to make the results from the disparate surveys as comparable to the \WD\ data set as possible. 
The complete list of surveys used is found in Table~\ref{tbl:dat:surveys}. 

\floattable
\begin{deluxetable}{lccccll}
	\tabletypesize{\scriptsize}
	\tablewidth{1.2\textwidth}
	\tablecaption{Summary of Spectroscopic Surveys Used}
	\tablehead{ \colhead{Survey} & \colhead{Release} & \colhead{Redshifts}  & \colhead{Coverage $(\Omega)$} &  \colhead{Band} & \colhead{$m_{\mathrm{lim}}$} & \colhead{Reference } \\
		  & \colhead{version} & \colhead{min/median/max}  &  \colhead{$(\mathrm{deg}^2)$} & & \colhead{AB mag} & }
	\startdata
		\href{http://www.6dfgs.net/}{6dFGS} & 3 & 0.01 / 0.05 / 0.20 & $1.37\times10^4$\tablenotemark{a} & $K_s$ & 11.25/14.49 & \cite{Jones:2009} \\
		\href{http://classic.sdss.org/dr7/}{SDSS-DR7} & 7 & 0.01 / 0.10 / 0.33 & $7.88\times10^3$\tablenotemark{b} & $r$ & 13.0/17.77 & \cite{SDSSdr7} \\
		\href{http://www.gama-survey.org/}{GAMA} & 2 & 0.01 / 0.18 / 0.43 & $144$ & $r$ & 14.0/19.0 & \cite{Baldry:2010} \\
		\href{http://iopscience.iop.org/article/10.1088/0067-0049/200/1/8/meta}{AGES} & 1 & 0.05 / 0.31 / 1.00 & $7.75$ & $I$ & 15.5/18.9/20.4 & \cite{AGES}\\
		\href{http://www.eso.org/sci/observing/phase3/data_releases.html\#other_programmes}{zCOSMOS-10k} & 2 & 0.05 / 0.61 / 1.00 & $1.7$ & $I^*$& 15.0/22.5 & \cite{Lilly:2009} \\
		\WD & 2 & 0.05 / 0.38 / 1.00 & $0.190$ & W1 & 15.0/18.70/19.14\tablenotemark{c} & \cite{Lake:2012} \\
	\enddata
	\tablecomments{Redshift surveys used to construct the samples here. 
	The selection for zCOSMOS was done using the Hubble Space Telescope's Advanced Camera for Surveys (ACS) filter F814W, which is approximately $I$-band. 
	The selections for AGES and \WD are split into a complete bright and sparse faint samples. 
	The Redshifts column contains the median redshift of the survey, and the minimum and maximum redshifts likely to be useful. 
	The smaller surveys, for example, have a bias against selecting galaxies that are local, large, and resolved because they would obstruct the field of higher redshift galaxies, so they require a higher minimum redshift cutoff. 
	The larger surveys, contrastingly, when they contain high redshift sources they are more likely to be redshift blunders, and so they require a lower maximum redshift. }
	\tablenotetext{a}{ Inital 6dFGS area is $1.7\times10^4\operatorname{deg}^2$, but all data with $\delta \ge -11.5\deg$ was removed to eliminate overlap with other surveys.} 
	\tablenotetext{b}{ Initial SDSS area is $8.04\times10^3\operatorname{deg}^2$, but the footprint of the smaller, deeper, surveys also used here are removed to eliminate overlap. } 
	\tablenotetext{c}{ The upper limit is in $R$-band magnitudes, as required in the Keck/DEIMOS documentation, and measured in the USNO's NOMAD catalog \citep{NOMAD}.}
	\label{tbl:dat:surveys}
\end{deluxetable}

\floattable
\begin{deluxetable}{l|lcc}
	\tabletypesize{\scriptsize}
	\tablewidth{0.86\textwidth}
	\tablecaption{Photometric Surveys Used by Spectroscopic Survey}
	\tablehead{ \colhead{Spectroscopic Survey} & \colhead{Photometric Survey} & \colhead{Bands} & \colhead{Citation } }
	\startdata
		\WD\tablenotemark{a} & \href{http://galex.stsci.edu/GR6/}{GALEX gr7} & FUV, NUV & \cite{Martin:2005} \\
		& \href{http://www.ipac.caltech.edu/2mass/}{2MASS} & $J$, $H$, $K_s$ & \cite{2mass} \\
		& \href{http://www.sdss3.org/dr10/}{SDSS-DR10} & $u$, $g$, $r$, $i$, $z$ & \cite{Ahn:2014} \\
		& \href{http://irsa.ipac.caltech.edu/Missions/wise.html}{AllWISE} & W1, W2, W3, W4 & \cite{Wright:2010} \\
		\hline
		6dFGS & \href{http://galex.stsci.edu/GR6/}{GALEX gr7} & FUV, NUV & \cite{Martin:2005} \\
		& \href{http://www.ipac.caltech.edu/2mass/}{2MASS} & $J$, $H$, $K_s$ & \cite{2mass} \\
		& \href{http://irsa.ipac.caltech.edu/Missions/wise.html}{AllWISE} & W1, W2, W3, W4 & \cite{Wright:2010} \\
		\hline
		SDSS & \href{http://www.sdss3.org/dr10/}{SDSS-DR10} & $u$, $g$, $r$, $i$, $z$ & \cite{SDSSdr10} \\
		& \href{http://www.ipac.caltech.edu/2mass/}{2MASS} & $J$, $H$, $K_s$ & \cite{2mass} \\
		& \href{http://irsa.ipac.caltech.edu/Missions/wise.html}{AllWISE} & W1, W2, W3, W4 & \cite{Wright:2010} \\
		\hline
		GAMA & \href{http://www.gama-survey.org/}{GAMA} & FUV, NUV & \cite{Liske:2015} \\
		& \href{http://classic.sdss.org/dr7/}{SDSS-DR7} & $u$, $g$, $r$, $i$, $z$ & \cite{SDSSdr7} \\
		& \href{http://www.gama-survey.org/}{UKIDSS LAS} & $Y$, $J$, $H$, $K$ & \cite{Lawrence:2007} \\
		& \href{http://irsa.ipac.caltech.edu/Missions/wise.html}{AllWISE} & W1, W2, W3, W4 & \cite{Wright:2010} \\
		\hline
		AGES & \href{http://www.sdss3.org/dr10/}{SDSS-DR10} & $u$, $g$, $r$, $i$, $z$ & \cite{SDSSdr10} \\
		& \href{http://archive.noao.edu/ndwfs/}{NDWFS-DR3} & $B_w$, $R$, $I$, $K$ & \cite{NDWFS} \\
		& \href{http://irsa.ipac.caltech.edu/data/SPITZER/SDWFS/}{SDWFS-DR1.1} & c1, c2, c3, c4 & \cite{Ashby:2009} \\
		& \href{http://irsa.ipac.caltech.edu/Missions/wise.html}{AllWISE} & W1, W2, W3, W4 & \cite{Wright:2010} \\
		\hline
		zCOSMOS & \href{http://irsa.ipac.caltech.edu/Missions/cosmos.html}{COSMOS} & FUV, NUV, $u^*$, $B_j$, $g^+$, $V_j$, &  \\
		&  & $r^+$, F814W, $i^+$, $i^*$, $z^+$, $J$, $K_s$ &  \cite{Capak:2007} \\
		& \href{http://www.sdss3.org/dr10/}{SDSS-DR10} & $u$, $g$, $r$, $i$, $z$ & \cite{SDSSdr10} \\
		& \href{http://irsa.ipac.caltech.edu/Missions/cosmos.html}{S-COSMOS-DR3} & c1, c2, c3, c4 & \cite{Sanders:2007} \\
		& \href{http://irsa.ipac.caltech.edu/Missions/wise.html}{AllWISE} & W1, W2, W3, W4 & \cite{Wright:2010} \\
	\enddata
	\tablecomments{ Photometric surveys used for fitting SEDs to sources, in order of increasing wavelength. }
	\tablenotetext{a}{ Not all sources have all data available, whether it was a question of coverage or depth, so roughly half of the sources were only characterized by \WISE\ data. }
	\label{tbl:dat:photbysurvey}
\end{deluxetable}

We cross matched the surveys to the AllWISE catalog \citep{AllWISE} and reject table using a spatial match with a $6\arcsec$ radius, the full width at half maximum (FWHM) of the \WISE\ beam, keeping only the nearest matching source. 
We then traversed the list again to ensure that each source from AllWISE was associated with only one target from the redshift survey, assigning the AllWISE source to the closest target in cases where multiple targets matched a single source. 
The reason for this choice is that the target closest to the photo-center of the AllWISE source is likely the one providing the dominant contribution to the detected flux. 
While an ad-hoc deblending procedure could have been developed, the issue was infrequent enough to make addressing it in this fashion not worthwhile; less than 3\% of sources in zCOSMOS-10k, the deepest, and narrowest, survey in this paper. 

Performing the initial search out to $6\arcsec$ allowed us to examine how the match radius affected both the completeness and purity of the sample. 
In that analysis we decided that keeping only sources with a match within half of the \WISE\ beam's FWHM ($3\arcsec$) was, subjectively, an adequate compromise among all of the sample completeness and purity factors, since the target likely contributes the majority of the flux in the \WISE\ measurement. 
Sources with matches between $6\arcsec$ and $3\arcsec$ that pass all other tests are regarded as lost to contamination, and are treated as a reduction in completeness for the survey. 
The same is done for sources which are flagged as having contaminated photometry in W1 in the AllWISE database (\code{w1cc\_map} $\neq 0$). 
The fraction of points lost to contamination by these criteria are: 9.2\% for 6dFGS, 4.1\% for SDSS, 3.8\% for GAMA, 3.0\% for AGES, 11.8\% for zCOSMOS, and 1.4\% for \WD.

\begin{figure}[htb]
	\includegraphics[width=0.48\textwidth]{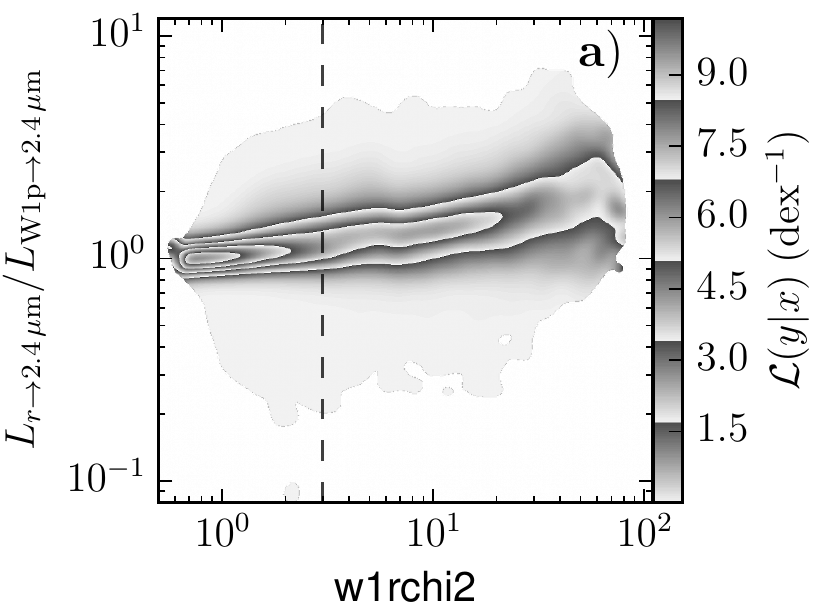}
	\includegraphics[width=0.48\textwidth]{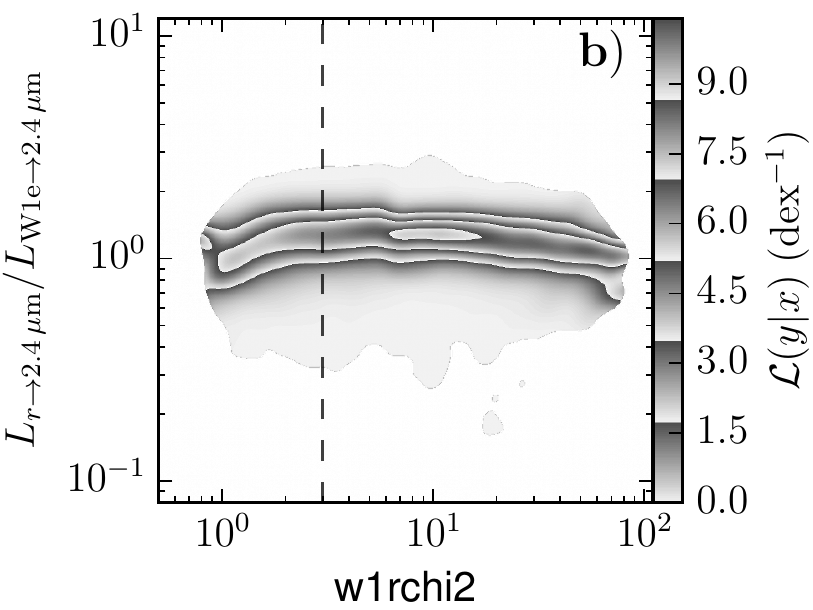}
		
	\caption{Plots of $L_{r\rightarrow2.4\micron} / L_{\mathrm{W1}\rightarrow2.4\micron}$ versus the reduced $\chi^2$ of the W1 PSF photometry fit. 
	$L_{r\rightarrow2.4\micron}$ is the $2.4\micron$ luminosity of the galaxy as predicted by $K$-correcting the SDSS profile fit photometry in $r$, and $L_{\mathrm{W1}\rightarrow2.4\micron}$ is the same prediction from $K$-correcting a W1 flux measurement. 
	Panel~\textbf{a} uses the PSF photometry for W1, and Panel~\textbf{b} uses the elliptical aperture fluxes. 
	The data used is described in Section~\ref{sec:sel:SDSSdets}, with the additional restriction that points have non-upper limit elliptical aperture fluxes. 
	The vertical line at reduced $\chi^2 = 3$ shows the separation we adopted between data for which the PSF flux was preferable and that for which the elliptical flux was preferable.}
	\label{fig:dat:psfvell}
\end{figure}

The large \WISE\ beam means that the vast majority of galaxies detected in the AllWISE data release are unresolved and well characterized by the point-spread function (PSF) photometry stored in the \code{w?flux} columns of the database. 
While there were not enough resources to perform a full and independent extended source analysis of the \WISE\ survey, the team was able to place elliptical apertures on sources that are already identified in the 2MASS Extended Source Catalog (XSC). 
Figure~\ref{fig:dat:psfvell} shows the trend in $L_\nu(2.4\micron)$ computed by $K$-correcting SDSS $r$ model fluxes (profile fit flux measurements) versus using \WISE\ W1 fluxes as a function of the reduced $\chi^2$ of the W1 PSF fit on sources from the SDSS survey. 
Panel \textbf{a} shows the trend when using W1 PSF photometry, and panel \textbf{b} shows the trend when using the W1 elliptical apertures. 
While both sets of data have a trend, the elliptical aperture has a smaller trend at large $\chi^2$. 

The hazard of using only PSF fluxes for resolved sources (even for only marg\-in\-al\-ly resolved ones) is twofold: first, $L_\star$, the luminosity scale at which there is a ``knee'' in the luminosity function, will be underestimated for low redshift galaxies; leading to the second, that the evolution in $L_\star$ will be overestimated. 
This bias will have further effects on the values observed in other LF parameters. 
$\kappa_\star$, the normalization of flux counts, should be slightly decreased because the bias is a blunting of the flux counts histogram at the bright end. 
The definition of $\phi_\star$, the value of the luminosity function at $L_\star$, makes its value dependent on $L_\star$, therefore both its value and the evolution of that value will be strongly affected. 
We therefore attempt to minimize this bias by using the elliptical aperture flux when \code{w?rchi2} $\ge 3$, if it is not an upper limit and if the source is within $5\arcsec$ of the XSC source (\code{xscprox} $\le5$). 

In the future, the ideal solution would be to generalize the photometry software used in producing the AllWISE catalog, WPHOT\footnote{\url{http://wise2.ipac.caltech.edu/docs/release/allsky/expsup/sec4_4c.html}}, to process the images used to generate each survey's target list at the same time as the \WISE\ images using a profile model for the galaxies; similar to what was done in \cite{Lang:2014}. 
We have intentionally chosen not to use \cite{Lang:2014} for the present work because it would introduce an additional systematic difference between the sources for which SDSS photometry is available, and those for which it is not, and we want to minimize such differences wherever possible.

\begin{figure*}[p]
	\includegraphics[width=\textwidth]{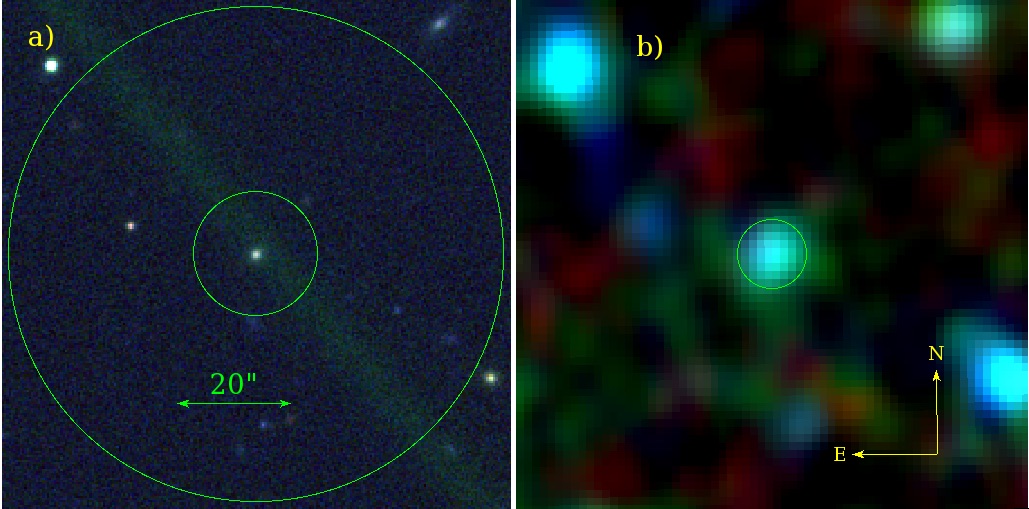}
	\caption{Panels show a source with contaminated optical photometry that causes a source that should not have been selected for the redshift sample to be included. 
	Panel~\textbf{a} is centered on the target source with SDSS $gri$ mapped to blue, red, and green, respectively, in a linear scale with the scale of each channel set to fit the source. 
	The green streak is almost certainly spurious contamination from a foreground moving object like an airplane or meteor.
	The outer circle is the Petrosian photometry aperture, used to measure the flux in determining targeting, and the inner circle is the aperture that contains half of the flux of the outer one. 
	Panel~\textbf{b}  shows the same field, as imaged by \WISE, with W1 mapped to yellow, and W2 to red. 
	The green circle on this panel shows the twice the FWHM of the native \WISE\ W1 beam. 
	The coordinates of this source are: $138.33114\deg$, $19.70335\deg$ in J2000 right ascension and declination. }
	\label{fig:dat:contam}
\end{figure*}

\begin{figure*}[p]
	\includegraphics[width=\textwidth]{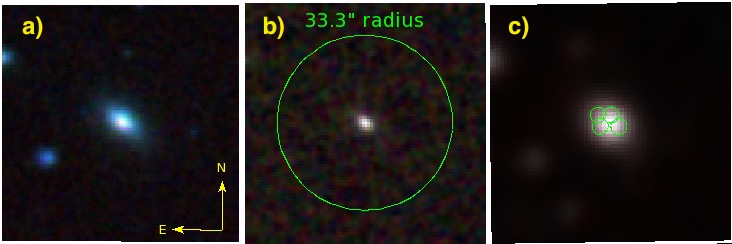}
	\caption{Panels show different views of a source that is large enough for \WISE\ PSF flux measurements to miss a significant portion of its flux. 
	This particular example happens to have been divided into four different sources by the WPHOT pipeline. 
	Panel~\textbf{a} shows the galaxy as seen in the Digital Sky Survey 2, with $B$, $R$, and $I$ mapped to blue, green, and red. 
	Panel~\textbf{b} shows the 2MASS view, with $J$, $H$, and $K_s$ mapped to blue, green, and red. 
	The $33.3\arcsec$ circle shows the radius out to which the radial profile was integrated to calculate the flux used in selection. 
	Panel~\textbf{c} shows the same galaxy as shown in the AllWISE coadd atlas, with W1 mapped to yellow, and W2 to red. 
	The green circles have $3.1\arcsec$ radii, and show the sources into which the galaxy is divided in the AllWISE database.  
	The coordinates of this source are: $24.48083\deg$, $-50.36086\deg$ in J2000 right ascension and declination. }
	\label{fig:dat:splitSrc}
\end{figure*}

Model fitting techniques that correct for model completeness using a smoothly varying selection function, as we intend to apply to this data set in the companion work LW17III, tend to make the model parameters particularly sensitive to outliers that are in low completeness regions. 
The most prominent example of this outlier effect in this work are objects which pass all of the selection flux cuts, but are in a position in the luminosity-redshift plane that has an extremely low selection probability assigned to it by the model. 
The two most common reasons for this are: sources with contaminated optical photometry; and sources that are marginally resolved by \WISE, but which do not have elliptical aperture photometry available.  
We calculated $L_\nu(2.4\micron)$ from the \WISE\ photometry and, if it isn't contaminated to the same extent as the optical selection photometry, then the source can be an outlier on the low side in a luminosity-redshift graph. 
Low luminosity outliers are in a region that the completeness model assigns a low probability of having been selected. 
This causes a large swing in the estimate of the LF model parameters in order to get a finite density at the position of the outlier. 
An example of a source with contaminated targeting photometry can be found in Figure~\ref{fig:dat:contam}, and one that lacks an elliptical aperture in AllWISE but the PSF flux is incorrect is in Figure~\ref{fig:dat:splitSrc}.
The apparent radius of a galaxy is correlated to both its luminosity and redshift, and the probability of significant contamination varies inversely with flux and with the area subtended on the sky by the source, leading to selection effects and biases. 
In principle these effects can be modeled if a radius-luminosity relationship is added onto the overall model, but that would require a set of radius measurements that is consistent across surveys, and that also has an accurate determination of the effect of seeing.

\begin{figure}[htb]
	\begin{center}
	\includegraphics[width=0.48\textwidth]{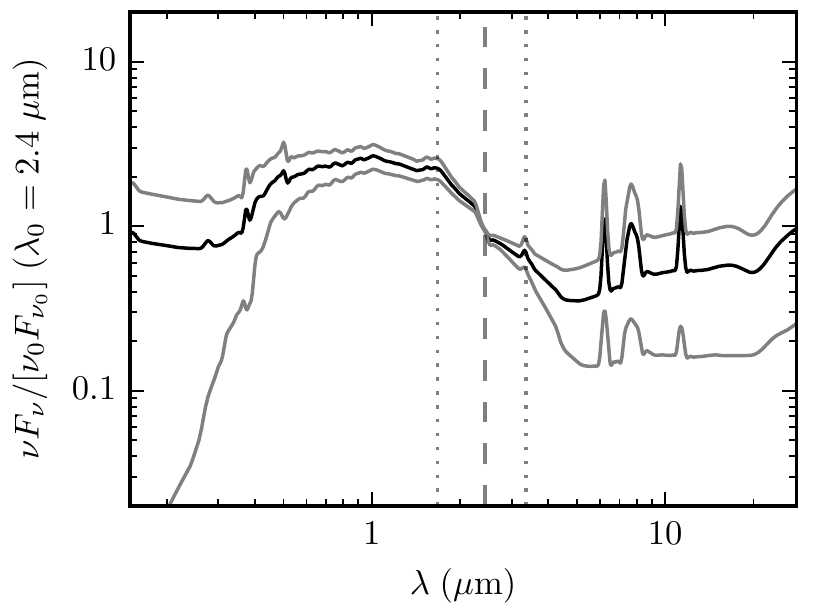}
	\end{center}
	
	\caption{ The mean SED of galaxies in the redshift range $0.05 < z \le 1$, as measured using the data and SED fits from \cite{Lake:2016}. 
	The solid grey lines show the 1-$\sigma$ type variance band around the mean SED, the dashed vertical line is at $2.4\micron$ and the vertical dotted lines show the effective wavelengths of the \WISE\ W1 filter for galaxies at the extreme redshifts of the sample, $z=0,\ 1$. }
	\label{fig:dat:meanSED}
\end{figure}

The factors discussed in the previous paragraph make it necessary to cut data with low selection probability and low luminosity as outliers. 
For that reason, a reduced maximum redshift was applied to 6dFGRS, SDSS, and GAMA, as shown in Table~\ref{tbl:dat:surveys}. 
Further, the minimum selection fluxes were extrapolated into luminosity cuts using the mean SED for all galaxies measured using the data from \cite{Lake:2016}. 
The fractions of $2.4\micron$ luminosity contributed by each of the templates from \cite{Assef:2010}, with the median AGN obscuration, can be found in Table~\ref{tbl:dat:meanpars} and a graph of the mean SED, with a 1-$\sigma$ type variance band around it, is in Figure~\ref{fig:dat:meanSED}. 
For AGES and \WD, surveys with a tiered target selection strategy, each survey was treated as though it were comprised of two fully independent surveys for this cut, with the division line set by the intermediate magnitude limit in Table~\ref{tbl:dat:surveys}. 
Finally, if the survey documentation did not explicitly cite a maximum flux limit, then one was imposed that cut the brightest few sources in order to ensure an accurate upper flux limit for the survey.

\begin{deluxetable}{lllll}
	\tabletypesize{\scriptsize}
	\tablewidth{0.54\textwidth}
	\tablecaption{Mean SED Parameters}
	\tablehead{ \colhead{$\langle f_{\mathrm{Ell}} \rangle$} & \colhead{$\langle f_{\mathrm{Sbc}} \rangle$} & \colhead{$\langle f_{\mathrm{Irr}} \rangle$} & \colhead{$\langle f_{\mathrm{AGN}} \rangle$} & \colhead{$\overline{\tau_B - \tau_V}$\tablenotemark{a}} }
	\startdata
		$0.490$ & $0.269$ & $0.114$ & $0.127$ & $0.023$ \\
	\enddata
	\tablecomments{ Mean of the $2.4\micron$ luminosity template fractions, alongside the median excess extinction on the AGN. 
	Numbers are given to three decimal places regardless of experimental uncertainty. }
	\tablenotetext{a}{ $\overline{\tau_B - \tau_V}$ here means the median of $\tau_B - \tau_V$.  }
	\label{tbl:dat:meanpars}
\end{deluxetable}

The \WISE\ All-Sky data release had a known and documented\footnote{\url{http://wise2.ipac.caltech.edu/docs/release/allsky/expsup/sec6_3c.html\#flux_under}} overestimation of the background, leading to an underestimation of the flux for faint sources. 
The AllWISE release remedied most, but not all, of the problem\footnote{\url{http://wise2.ipac.caltech.edu/docs/release/allwise/expsup/sec2_2.html}}. 
We, therefore, added a small flux to correct for the over-subtraction in the PSF photometry, on average, in W1 and W2. 
The values added are $1.5\microJy$ and $7\microJy$ in W1 and W2, respectively, as shown in Table~6 of the catalog completeness section of the AllWISE Explanatory Supplement\footnote{\url{http://wise2.ipac.caltech.edu/docs/release/allwise/expsup/sec2_4a.html}}. 
The aperture photometry was not affected by this issue, so when elliptical aperture photometry was used in this work it was unaltered.

%

\begin{figure}[htb]
	\begin{center}
	\includegraphics[width=0.48\textwidth]{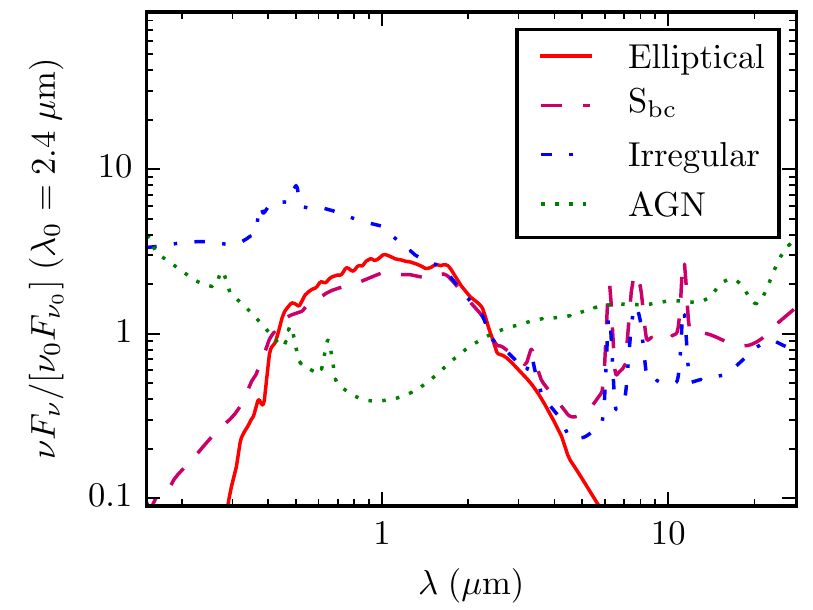}
	\end{center}
	
	\caption{The template spectra used from \cite{Assef:2010}. 
		The red solid line is the template called ``Elliptical." 
		The purple dashed line is ``Sbc." 
		The blue dash-dotted line is ``Irregular." 
		And the green dotted line is ``AGN," unobscured. }
	\label{fig:dat:templates}
\end{figure}

Targets from each survey were also matched to other photometric surveys using a $1\arcsec$ spatial match in order to obtain more photometric points to use in modeling the spectral energy distributions (SEDs) of the galaxies. 
The SED models were made using the four templates from \cite{Assef:2010}. 
The set of templates contained four basis galaxies, identified as Elliptical, Sbc, Irregular, and Active Galactic Nucleus (AGN). 
All of the models are normalized to be $10^{10}\Lsun$ in the wavelength range 0.03--$30\micron$. 
The AGN template, additionally, has a dust obscuration model parametrized by $\operatorname{E}(B-V)$. 
The models, normalized to unit luminosity at $2.4\micron$ with the AGN unobscured, can be found in Figure~\ref{fig:dat:templates}. 

The templates were constructed to be fit to fluxes by minimizing $\chi^2$ with respect to to a linear combination of the template fluxes with non-negative coefficients, and a search in the 1-dimensional parameter space for the best AGN extinction, $\operatorname{E}(B-V) = (2.5 / \ln(10)) \cdot (\tau_B - \tau_V)$. 
That is, the model has the form:
\begin{align}
	F_\nu(\nu, z) & = a_\mathrm{E} F_\mathrm{E}(\nu, z) + a_\mathrm{S} F_\mathrm{S}(\nu, z) \nonumber \\
		&\hphantom{=}+ a_\mathrm{I} F_\mathrm{I}(\nu, z) + a_\mathrm{A} F_\mathrm{A}(\nu, z, \tau_B - \tau_V), \label{eqn:sedform} \\
	\chi^2 & = \sum_{i \in \{\mathrm{filters}\}} \left(\frac{F_{\mathrm{obs}\,i} - F_{\mathrm{mod}\,i}}{\sigma_i}\right), \label{eqn:sedchisqr}
\end{align}
with all $a_i \ge 0$, and $12 > \tau_B - \tau_V \ge 0 $. 
The $a_i$ were fit using the SciPy \code{optimize} package's routine \code{nnls} (quadratic programming for non-negative least squares), and $\tau_B - \tau_V$ were fit with the routine \code{brent} (Brent's algorithm) with fallback to \code{fmin} (Nelder-Meade simplex).

There is one modification to that procedure for the fits done for this paper. 
The templates do not include the possibility of adjustable dust obscuration of the stellar population, so a dusty starburst that has a detection in \WISE's $12\micron$ filter, W3, will often be best fit with a galaxy that is dominated by its Elliptical component (to satisfy optical redness) and a super-obscured AGN ($\tau_B - \tau_V > 12$) masquerading as the emission from the stellar dust component. 
The problem this creates is that it makes the SED fit the data more poorly in the most important range for work on the $2.4\micron$ luminosity function, where K-corrections from W1 to $2.4\micron$ are performed. 
We used two techniques to work around this problem. 
First, we limited the excess in optical depth as $\tau_B - \tau_V \le 12$ (equivalently, $\operatorname{E}(B-V) \le 13.03$). 
Second, when the SED was badly modeled ($\chi^2 > \mathrm{max}(N_{\mathrm{df}}, 1) \times 100$) and unlikely to be an AGN ($W1 - W2 > 0.5\operatorname{Vega\ mag}$), we used the best model with an unobscured AGN, $\operatorname{E}(B-V) = 0$. 
The reduced $\chi^2$ criterion was determined by eye, and the AGN selection was found in \cite{Assef:2013} to select low redshift AGN with 90\% completeness.

Limiting the excess optical depth, $\tau_B - \tau_V$, to be non-negative introduces a bias to the parameter estimation of the individual galaxies. 
It is even physically possible for a source to appear bluer than expected if the line of sight is unobscured and dust clouds are reflecting excess blue light into it (that is, the line of sight contains a significant contribution from reflection nebulae in the target galaxy). 
Even so, applying a negative optical depth excess to dust obscuration models is not likely to produce an accurate spectrum for reflection (since obscuration models both reflection and absorption), and the magnitude of the negative excess doesn't have to be large to cause the estimate of the maximum redshift at which the galaxy would be included in the sample to diverge, outweighing the impact of biases introduced by requiring $\tau_B-\tau_V$ to be non-negative.

Admittedly, the limitation that the $a_i$ be non-negative introduces a source of potential bias to analyses done using them, but for their intended use, predicting SEDs at unobserved wavelengths, allowing components to take negative values produces unphysical outliers of the same sort that allowing $\tau_B-\tau_V < 0$ does, and not an insignificant number of them. 
While the components in Figure~\ref{fig:dat:templates} all look very different, and therefore unlikely to produce large negative coefficients in the fit, it should be kept in mind that the graph has a logarithmic scale in flux and the fitting is done linearly in flux. 
Thus, depending on what rest frame wavelengths we have photometry of the target galaxy at, and the signal to noise ratio of the photometry, the Elliptical and Sbc templates are similar (about $0.5$--$5\micron$), or the Sbc and Irregular templates are similar (about $2$--$5\micron$).
Experimentally, we examined the resulting $a_i$ when using ordinary least squares (OLS) in the zCOSMOS data set, where we have the largest amount of auxiliary data and should, therefore, expect the best behavior.
Nearly all sources in the sample had at least one negative $a_i$, and about a quarter of sources had a an $a_i$ that was more negative than half the sum of the $a_i$ from doing the non-negative version of the fit.
For some of them, the sum of the OLS $a_i$, which should correspond to the overall unobscured luminosity of the galaxy, was outright negative.

The AllWISE data release also has an issue where the flux uncertainties in W1 could be overestimated, or even missing, in the ecliptic longitudes covered by the ``3-band cryo" portion of the survey\footnote{\url{http://wise2.ipac.caltech.edu/docs/release/allwise/expsup/sec2_2.html\#w1sat}}. 
The flux uncertainties are needed to model SEDs using Equation \ref{eqn:sedchisqr}, so we substitute an uncertainty calculated from \code{w1sigp2}, the uncertainty in the mean flux measured from individual calibrated frames by the \WISE\ photometry system, in magnitudes. 
This quantity differs from, and is usually less reliable than, the standard flux and uncertainty columns.  
This is because the standard uncertainty is calculated by simultaneously fitting all of the frames at the same time, and \code{w1sigp2} is calculated measuring the flux on each frame individually. 
Therefore the standard columns are to be favored if there isn't strong evidence that the standard flux uncertainty is overestimated. 
Empirically, the substitution was justified when the following equation is satisfied:
\begin{align}
	\sigma_{\mathrm{W1}} > 2\sqrt{\left(0.02 F_{\mathrm{W1}}\right)^2 + \sigma_{\mathrm{W1p2}}^2},
\end{align}
with $\sigma_{\mathrm{W1p2}} \equiv 0.4 \ln(10) [\code{w1sigp2}] F_{0\,\mathrm{W1}} 10^{-0.4 [\code{w1magp}]}$. 
The relationship between \code{w1magp}, the mean flux (in magnitudes) for which \code{w1sigp2} is the uncertainty, and \code{w1sigp2} is more tightly correlated in the same way that the standard W1 flux and uncertainty columns are, making the use of \code{w1magp} in the conversion from magnitude to flux uncertainties preferable.

The template models were used to generate $K$-corrections from W1 flux to $2.4\micron$ rest frame luminosities using the equations from \cite{Hogg:2002} and \cite{Blanton:2003K}. 
We corrected to $2.4\micron$ rest frame luminosity in order to minimize the errors associated with K-correction for the overall sample in the same fashion as was done in \cite{Blanton:2003LF}. 
In other words, W1 fluxes were $K$-corrected to the wavelength W1 samples at the median redshift of sources with $F_{\mathrm{W1}} \ge 80 \microJy$ from \WD, $z=0.38$.


Details of how each survey was processed that are peculiar to each survey, and what auxiliary photometric data was used, can be found in the following subsections, starting with this work's defining survey, \WD, and then in decreasing order in survey area on the sky.

\subsection{\WD\ Details}
\WD\ consisted of a one night survey performed on the Keck II telescope using the DEIMOS instrument \citep{Faber:2003}, with resulting data reduced using the DEEP2 spec2d pipeline \citep{Newman:2013, Cooper:2012}, and analyzed using SpecPro \citep{Masters:2011}. 
\WD\  included observations of 10 different slit masks at disparate positions with high galactic latitude ($b > 30^\circ$). 
So, while the net area covered by those 10 masks is small, $5.78\times 10^{-5}\sr = 0.190 \operatorname{deg}^2$, the sample is less affected by cosmic variance than one might naively expect because the fields are non-contiguous. 
Though we do not estimate cosmic variance here directly, estimates of cosmic variance for larger surveys, for example \cite{Driver:2010} and \cite{Driver:2011}, suggest that it is not small compared to the shot noise level for $222$ sources, $(222)^{-1/2} \approx 6.7\%$. 
Because the source density varies with galactic latitude, the targeting completeness varies from field to field, necessitating the use of a selection function that varies by field.

A small number of sources, about 5, in \cite{Lake:2012} had incorrectly measured redshifts, or lack thereof. 
This is based on a closer reanalysis of the data with more consistent standards for when a redshift is to be assigned, as will be explained below in the discussion of quality codes. 
Tables~\ref{tbl:dat:WDdata1}~and~\ref{tbl:dat:WDdata2} contain a short excerpt from the machine readable table published along with this work. 
This table contains both more rows and more columns than the one published with \cite{Lake:2012}.

\floattable
\begin{deluxetable}{lrccccrrccc}
	\tabletypesize{\scriptsize}
	\rotate
	\tablewidth{0.86\textheight}
	\tablecaption{\WD\ Redshift Catalog DR2, Excerpt Part 1}
	\tablehead{ \colhead{\code{Designation}} & \colhead{\code{ID}} & \colhead{\code{Ra}} & \colhead{\code{Dec}} &\colhead{\code{W1targ}} & \colhead{\code{TG}} & \colhead{\code{mask}} & \colhead{\code{SlitNum}} & \colhead{\code{z}} & \colhead{\code{z\_err}} & \colhead{\code{q\_z}}  \\
	\colhead{} & & \colhead{\deg} & \colhead{\deg}& \colhead{Vega mag} & & & & & & }
	\startdata
		WISEPC J230334.10+040532.7 & 342 & 345.8921204 & 4.0924325 & 11.483 & 2 & 5 & 2 & 0.0 & nan & 4  \\
		WISEPC J230256.33+040511.2 & 351 & 345.7347412 & 4.0864501 & 13.531 & 2 & 5 & 11 & 0.0638 & 0.00017 & 4  \\
		WISEPC J230310.28+040517.3 & 352 & 345.7928467 & 4.088161 & 13.938 & 2 & 5 & 12 & 0.1862 & 0.000221 & 4  \\
		WISEPC J230325.34+040520.5 & 353 & 345.8556213 & 4.0890427 & 15.178 & 2 & 5 & 13 & 0.1224 & 8.79e-05 & 4 \\
		WISEPC J230315.00+040527.7 & 355 & 345.8125 & 4.0910487 & 13.871 & 2 & 5 & 15 & 0.0 & nan & 4 \\
		WISEPC J230336.45+040436.2 & 398 & 345.901886 & 4.076745 & 17.018 & 4 & 5 & 56 & 1.7771 & 0.00335 & 4  \\
		WISEPC J230257.17+040548.2 & 405 & 345.7382202 & 4.0967259 & 17.928 & 4 & 5 & 62 & nan & nan & -1 \\
	\enddata
	\tablecomments{Excerpt from the first set of columns from the machine readable table formatted data published with this paper. 
	\code{Designation} has ``NA" for serendipitous sources. 
	\code{ID} is a unique integer assigned to each source in the catalog. 
	\code{Ra} and \code{Dec} are the J2000 right ascension and declination of the primary target on the slit. 
	\code{W1targ} is the W1 Vega magnitude used for target selection (``nan" if unavailable). 
	\code{TG} is the target group (explained in the text). 
	\code{mask} is the mask number, and it corresponds to the Field Number of Table~\ref{tbl:dat:WDfields}. 
	\code{SlitNum} is the slit number in the mask the source fell on. 
	\code{z} is the Earth-centric redshift of the source (``nan" if no valid redshift could be determined); no correction for the motion of the sun, Earth's orbital motion, or the CMB dipole was made, but all redshifts were gathered on a single night, Universal Time 2010 September 14, using the Keck 2 telescope on Mauna Kea, Hawaii. 
	\code{z\_err} is the uncertainty in the redshift, as ascertained by the template correlation performed by SpecPro (``nan" if redshift invalid or a star). 
	\code{q\_z} is the quality code of the spectrum, explained in the text.}
	\label{tbl:dat:WDdata1}
\end{deluxetable}

\floattable
\begin{deluxetable}{l|rrccl}
	\tabletypesize{\scriptsize}
	\rotate
	\tablewidth{0.86\textheight}
	\tablecaption{\WD\ Redshift Catalog DR2 Excerpt, Part 2}
	\tablehead{ \colhead{\code{Designation}} & \colhead{\code{class}} &\colhead{\code{SpecProTemplate}} &  \colhead{\code{R}} & \colhead{\code{cont}} & \colhead{\code{SpecFeatures}} }
	\startdata
		WISEPC J230334.10+040532.7 & Star & M Star & 1 & 1 & NaI TiO BaI Ha TiO \\
		WISEPC J230256.33+040511.2 & Gal & Red Galaxy & 1 & 1 & MgI NaI BaI \\
		WISEPC J230310.28+040517.3 & Gal & Green Galaxy & 1 & 1 & G-band Hb MgI NaI NII Ha NII SII SII \\
		WISEPC J230325.34+040520.5 & Gal & Blue Galaxy & 1 & 1 & Hb NaI Ha NII SII SII \\
		WISEPC J230315.00+040527.7 & Star & G Star & 1 & 1 & MgI NaI BaI Ha \\
		WISEPC J230336.45+040436.2 & QSO & SDSS Quasar & 1 & 1 & AlIII CIII CII MgII \\
		WISEPC J230257.17+040548.2 & Unseen & NA & 0 & 0 & \\
	\enddata
	\tablecomments{Excerpt from the second set of columns (\code{Designation} repeated) from the machine readable table formatted data published with this paper. 
	\code{class} is the spectroscopic classification assigned to the source, one of: ``Star" for stars, ``Gal" for galaxy, ``QSO" for broad line quasar, ``Indet" for a source that had an indeterminate spectrum, ``Unseen" for sources that didn't produce an observable spectrum, and ``Lost" for sources that were lost to instrument constraints. 
	\code{SpecProTemplate} is the name of the SpecPro template that matches the spectrum closest (``NA" for invalid sources). 
	\code{R} is a flag for whether the source was ``real," that is, it corresponds to a non-artifact AllWISE source. 
	\code{SpecFeatures} is a list of spectroscopic features listed in the \code{specpro} software that were identified, in increasing wavelength order (no distinction is made between emission and absorption features). }
	\label{tbl:dat:WDdata2}
\end{deluxetable}

The reanalyzed data contains four main columns relevant for selecting subsamples. 
The column \code{TG}, short for `Target Group,' contains an integer encoding which group of targets the source was in. 
The values \code{TG} takes are: 1 for the central source of the DEIMOS slit mask, 2 for W1 bright sources ($F_{\mathrm{W1}} \ge 120\microJy$, in the \WISE\ Preliminary Release), 3 for W1 intermediate sources ($120\microJy > F_{\mathrm{W1}} \ge 80\microJy$), 4 for W1 faint or non-detected sources ($80 \microJy > F_{\mathrm{W1}}$), and 5 for targets that serendipitously fell on the slit of a target. 
For analysis of pseudo-randomly selected galaxies with well known selection completeness, targets with $1 < \code{TG} < 5$ should be used. 
In order to have good completeness of the initial detections we recommend further limiting the sample to $\code{TG} < 4$, as is done in LW17III.

The quality of the redshift is encoded in a column of integers named \code{q\_z}, and takes on the values: $-1$ for sources with no detected flux in the spectrum, 0 for sources that have a spectrum but for which it was not possible to even estimate a redshift, 1 for targets where a redshift measurement was possible but no spectral features could be identified (blunders could not be ruled out, confidence $ < 50\%$), 2 for targets where the redshift is better but still uncertain (confidence $<95\%$), 3 for targets that have a secure redshift with at least one clearly identifiable spectral feature or more of lesser quality (absorption or emission lines), and 4 for targets with multiple clearly identifiable spectral features. 

The analysis of the spectra allowed the targets to be broken up by classification, \code{class}. 
\code{class} takes on four possible values: ``Gal" for ordinary galaxies, ``QSO" for broad-line AGN, ``Star" for stellar spectra, ``Indet" for spectra of indeterminate type, ``Unseen" for sources without any detectable flux in the spectrum, and ``Lost" for sources lost to instrument constraints. 
Naturally, an analysis of extragalactic targets must be limited to ``Gal" and ``QSO" targets.

The last selection relevant column is \code{R}. 
\code{R} stands for ``Real" and takes the value 1 if the source produced a spectrum or can be associated with a non-artifact source in the AllWISE database, and 0 otherwise. 
Only targets with $\code{R} = 1$ are relevant.

The completeness of the W1 faint sample is much lower and more poorly defined compared to the brighter two, as can be seen by comparing the spectroscopy completenesses ($f_{Q\ge3}$) in Table~\ref{tbl:dat:WDfields}, so the combined sample defined in this work is limited to only targets with $F_{\mathrm{W1}} \ge 80 \microJy$ for all surveys. 
This was done in order to make the results from all the surveys as comparable as possible. 
The following subsections contain plots showing the  distribution of primary selection flux of the survey versus $F_{\mathrm{W1}}$. 
They show that the effect of both cuts must be accounted for when analyzing all surveys deeper than SDSS.

\floattable
\begin{deluxetable}{cccc}
	\tablewidth{0.75\textwidth}
	\tablecaption{\WD\ Field Completenesses}
	\tablehead{\colhead{Field} & \colhead{$\{\ge 120\}$} & \colhead{$\{80$--$120\}$} & \colhead{$\{<80\}$}  \\
	\colhead{Number} & \colhead{$(f_{Q \ge 3}/f_{\mathrm{targ}}/N_{\mathrm{tot}})$} & \colhead{$(f_{Q \ge 3}/f_{\mathrm{targ}}/N_{\mathrm{tot}})$} & \colhead{$(f_{Q \ge 3}/f_{\mathrm{targ}}/N_{\mathrm{tot}})$} }
	\startdata
	$1$  & $0.97/0.70/84$ & $1.00/0.29/34$ & $0.78/0.21/41$ \\  
	$2$ & $1.00/0.59/99$ & $0.90/0.30/33$ & $0.71/0.18/40$ \\  
	$3$ & $0.98/0.77/74$ & $0.92/0.37/38$ & $0.83/0.29/21$ \\  
	$4$ & $1.00/0.72/68$ & $0.55/0.42/26$ & $0.81/0.27/59$ \\  
	$5$ & $1.00/0.69/55$ & $0.89/0.36/25$ & $0.56/0.34/95$ \\  
	$6$ & $0.91/0.81/54$ & $0.85/0.38/34$ & $0.58/0.30/88$  \\  
	$7$ & $1.00/0.77/44$ & $1.00/0.52/15$ & $0.60/0.24/25$ \\  
	$8$ & $1.00/0.80/45$ & $0.91/0.46/24$ & $0.62/0.51/57$ \\  
	$9$ & $1.00/0.76/46$ & $0.75/0.51/39$ & $0.56/0.24/75$  \\  
	$10$ & $0.98/0.74/58$ & $0.72/0.49/37$ & $0.65/0.35/57$ \\
	\enddata
	\tablecomments{Spectroscopic and targeting completeness of \WD, broken down by field. 
	Specifically, the column contain the fraction of slits that produced high quality spectra / fraction of targets assigned slits / total available targets, broken down by W1 flux sample (limits in $\microJy$). 
	This table is adapted from Table~2 in \cite{Lake:2012} with an updated analysis of the spectra and target source types based on the AllWISE data release.}
	\label{tbl:dat:WDfields}
\end{deluxetable}

\begin{figure}[htb]
	\includegraphics[width=0.48\textwidth]{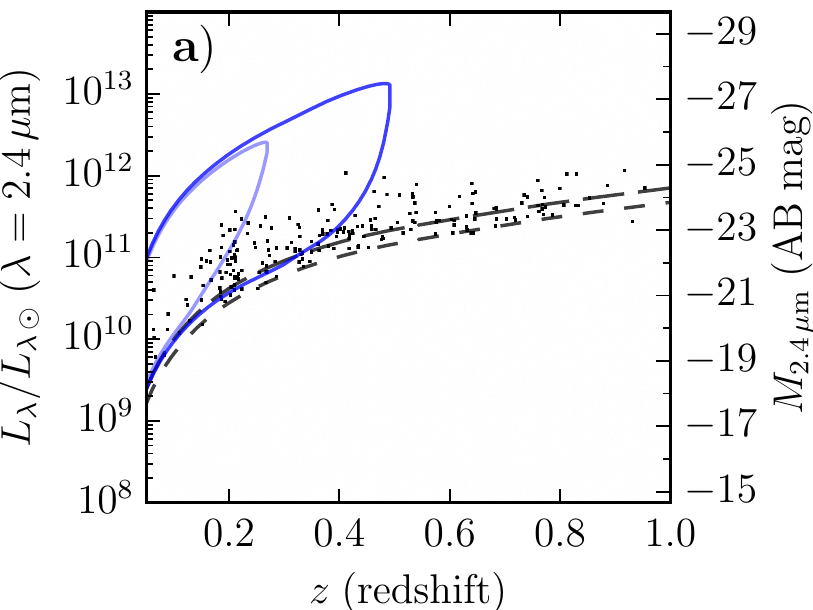}
	\includegraphics[width=0.48\textwidth]{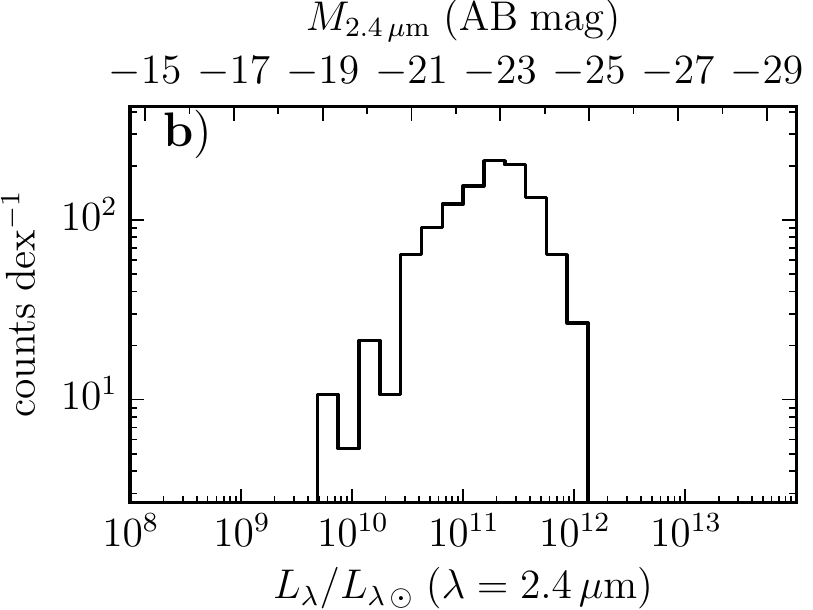}
	\includegraphics[width=0.48\textwidth]{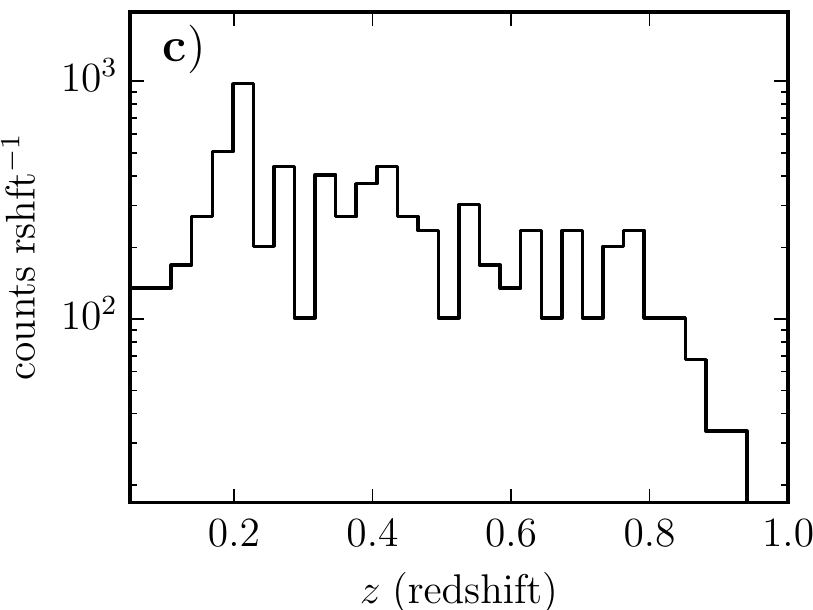}
	
	\caption{Panel \textbf{a} contains a scatter plot showing the range of luminosities and redshifts sampled by the \WD\ survey. 
	The dashed lines show luminosity cuts based on the mean SED from \cite{Lake:2016}: any source in the bright sample that falls below the long dashed line is cut, and any source in the faint sample that falls below the dashed line is cut. 
	The blue translucent lines show where the SED variety completeness is 95\% and 98\%, in order of increasing lightness.
	Panels \textbf{b} and \textbf{c} contain histograms of the data in redshift and luminosity, respectively, after the cuts in panel \textbf{a} are applied. 
	No completeness corrections are applied, and the distributions are normalized to the number of data points the set contributes, $N=222$ in \textbf{a} and $208$ in all other panels.}
	\label{fig:dat:zLumaWD}
\end{figure}

Figure~\ref{fig:dat:zLumaWD} is a scatter plot of redshifts versus luminosity, alongside the marginal histograms in redshift and log-luminosity for the sources used in the sample defined here. 
The plots are meant to show the raw quantity of data available at each redshift and luminosity, and thus contain no completeness corrections, and are normalized to the total number of data points. 
Of particular note, \WD\ contains few redshifts $z>1$, and only one with $z \le 0.05$. 
Given that the slit mask targeting avoided large resolved galaxies, the survey has a selection bias against redshifts lower than this, so we have limited the sources included from all surveys to be both low redshift ($z\le 1$), and have $z > 0.05$ for small area surveys or $z > 0.01$ for large area ones (6dFGS, SDSS, and GAMA).

Figure~\ref{fig:dat:zLumaWD}, panel~\textbf{a}, also contains blue curves bounding regions where the color variety completeness is approximately constant (to within $2\%$ for the light blue curve, and $5\%$ for the dark blue curve). 
Color variety completeness is defined by:
\begin{align}
	S_{\mathrm{color}}(L, z) \equiv \frac{\left\langle S(F_{\mathrm{sel}}, F_0, \vec{x}) \right\rangle_{\mathcal{L}_{\mathrm{SED}}} }{\operatorname{max}(S(F_{\mathrm{sel}}, F_0, \vec{x})) } , \label{eqn:dat:colvarcomp}
\end{align}
where $S(F_{\mathrm{sel}}, F_0, \vec{x})$ is the selection probability (completeness) for a galaxy at real space position $\vec{x}$ (for example, $\alpha$, $\delta$, and $z$) with two observer frame fluxes at different wavelengths, $F_{\mathrm{sel}}$ and $F_0$. 
The average in the numerator is weighted by the likelihood that a galaxy at redshift, $z$ and with (spectral) luminosity $L$ will be observed to have fluxes $F_{\mathrm{sel}}$ and $F_0$ (called $\mathcal{L}_{\mathrm{SED}}$, see Equations~19 of LW17I), including both color variety and a model for measurement noise. 
The denominator is the maximum value that $S(F_{\mathrm{sel}}, F_0, \vec{x})$ takes, removing factors like intentionally sparse sampling from $S_{\mathrm{color}}(L, z)$. 
The faint sample is automatically excluded from these regions because it is too narrow to provide a flat selection region.

The reason for including the blue curves in the plot is that they show the regions where the likelihood model defined in LW17I can be neglected. 
In the case of \WD, the $95\%$ curve leaves 91 sources ($43.8\%$ of the pre-cut data), and the $98\%$ curve leaves 40 ($19.2\%$ of pre-cut). 

\begin{figure}[htb]
	\begin{center}
	\includegraphics[width=0.48\textwidth]{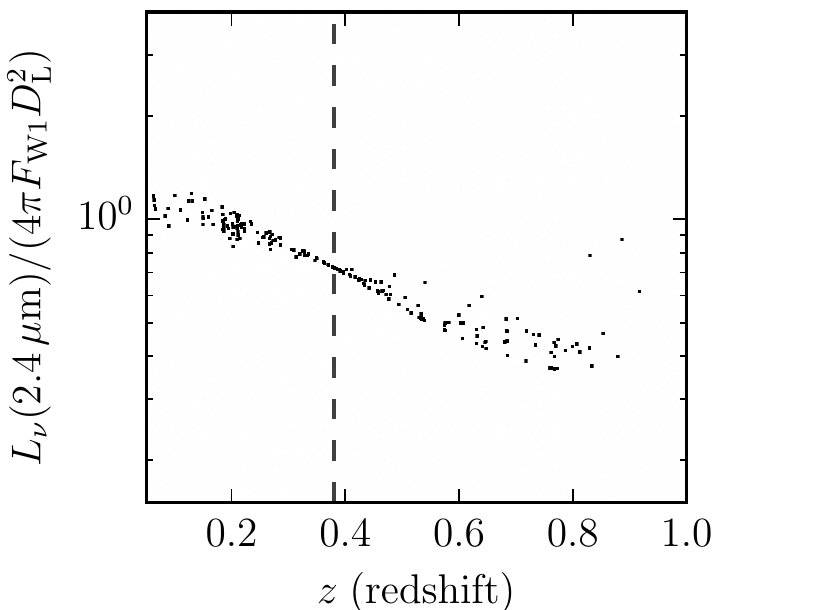}
	\end{center}
	
	\caption{Scatter plot of K-corrections from W1 to rest frame $2.4\micron$ applied to \WD\ data. 
	The vertical dashed line at $z=0.38$ marks the fiducial redshift used in this work, where the K-correction is only the bandwidth scaling by $1+z$.}
	\label{fig:dat:KcorrWD}
\end{figure}

The photometry from outside sources available for \WD\ was non-uniform, as mentioned in Table~\ref{tbl:dat:photbysurvey}. 
In total, roughly half of the sources have some photometry outside of AllWISE available, but that leaves only W1 and W2 photometry for the majority of the other half. 
This is not a problem for the accuracy of the $K$-corrections used in this LW17III, shown in Figure~\ref{fig:dat:KcorrWD}, because non-AGN galaxy SEDs are remarkably uniform in the wavelength range of interest ($1.7$--$3.4\micron$, see the $1$-$\sigma$ varience band around the mean in Figure~\ref{fig:dat:meanSED}). 
This is why \cite{Assef:2013} were able to show that this one color is remarkably good at picking out low redshift AGN, and therefore sufficient for narrowing the SED model in the wavelength range of interest.  
This fact also makes it extremely difficult to photometrically split the galaxies into red and blue types, as is done for most works on the galaxy luminosity functions. 
This is a problem because red cluster member galaxies typically have a different LF than bluer field galaxies, not to mention AGN. 
For that reason, most studies of the LF will remove AGN entirely, and perform two analyses on the ordinary galaxy data: one analysis with a single luminosity function, and one where the red and blue galaxies are modeled separately. 
The \WD\ data could be split by spectroscopic characteristics, but only broadly (for example, by the presence of emission lines), and performing a comparable analysis on the other surveys would have been prohibitively time consuming.

\subsection{6dFGS Details}
\begin{figure}[htb]
	\begin{center}
	\includegraphics[width=0.48\textwidth]{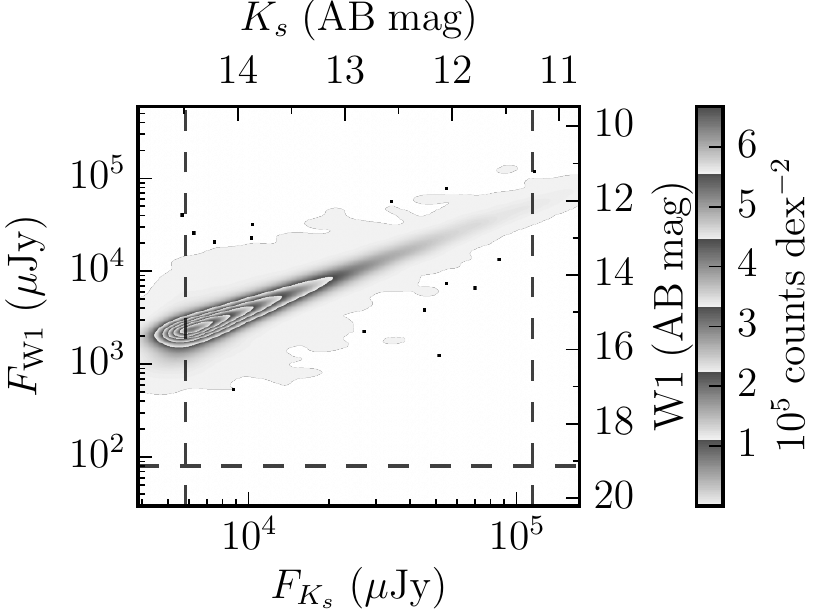}
	\end{center}
	
	\caption{Density plot showing the measured fluxes in relationship to the selection limits imposed by the 6dFGS survey ($14.49 \ge K_s > 11.25$) and here ($F_{\mathrm{W1}} \ge 80\microJy$). 
	Note how the additional \WISE\ limit eliminates no data.}
	\label{fig:dat:fluxflux6d}
\end{figure}

The 6dF Galaxy Survey (6dFGS) was originally defined in \cite{Jones:2004}, and the final data release used in this paper is described in \cite{Jones:2009}. 
6dFGS contains several sub-samples selected using different techniques, but the sample of primary interest to this paper is the one selected from the Two Micron All Sky Survey (2MASS) extended source catalog using the $K_s$-band flux. 
This subset is designated as having $\code{PROGID} =  1$, and satisfies $K_s < 14.49 \operatorname{AB\ mag}$. 
The reason the $K_s$ selected sample is the most relevant to this work is because $K_s$ is adjacent to W1 in wavelength, and so subject to less potential selection bias than surveys that were selected optically, as the flux-flux graph in Figure~\ref{fig:dat:fluxflux6d} shows.

Selecting the subset of redshifts with high confidence is relatively straightforward with 6dFGS. 
Those targets with $3 \le \code{quality} < 6$, where `\code{quality}' is the name of the column of integers classifying redshifts by the quality (\code{QUALITY} in the 6dFGS schema), selects for targets with science quality redshifts ($\code{quality} \ge 3$) and removes those that are Milky Way sources (they have $\code{quality} = 6$). 

Further, the 6dFGS data set contains multiple redshift for a fraction of the sources.
When multiple redshifts are available for a source, the selection of which redshift to use involved a couple of steps. The primary discriminator is \code{quality}; when they differ the sort order, in decreasing preference, is: 4, 3, 6, 2, and 1. When multiple redshifts have the same quality code the redshift with the lower measured uncertainty, \code{ZFINALERR} in the 6dFGS schema, is preferred if one or more redshifts had measured uncertainties (note well: a value of $0$ in the uncertainty column is not measured).

\begin{figure}[htb]
	\includegraphics[width=0.48\textwidth]{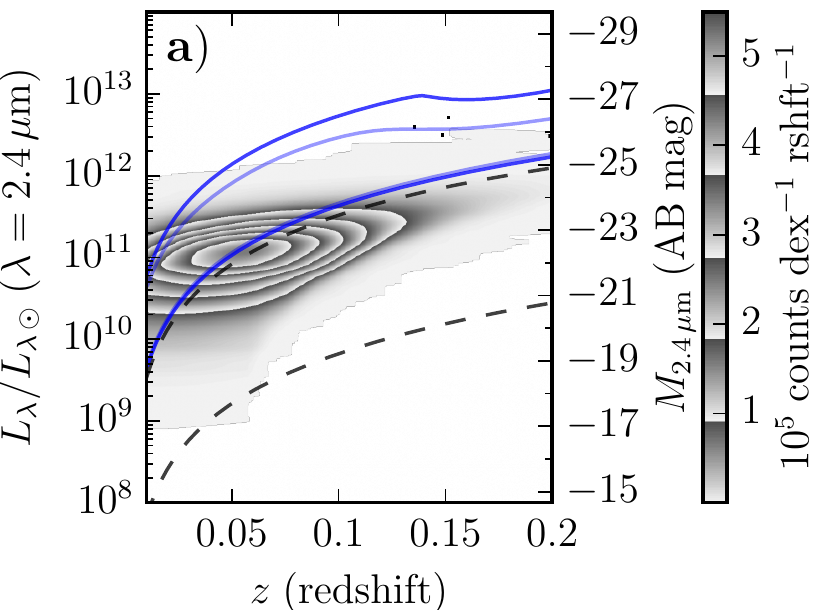}
	\includegraphics[width=0.48\textwidth]{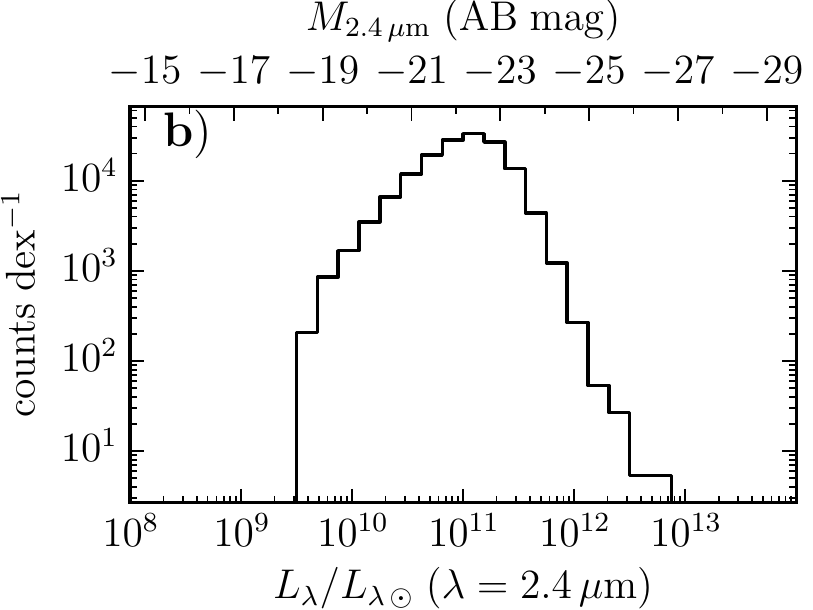}
	\includegraphics[width=0.48\textwidth]{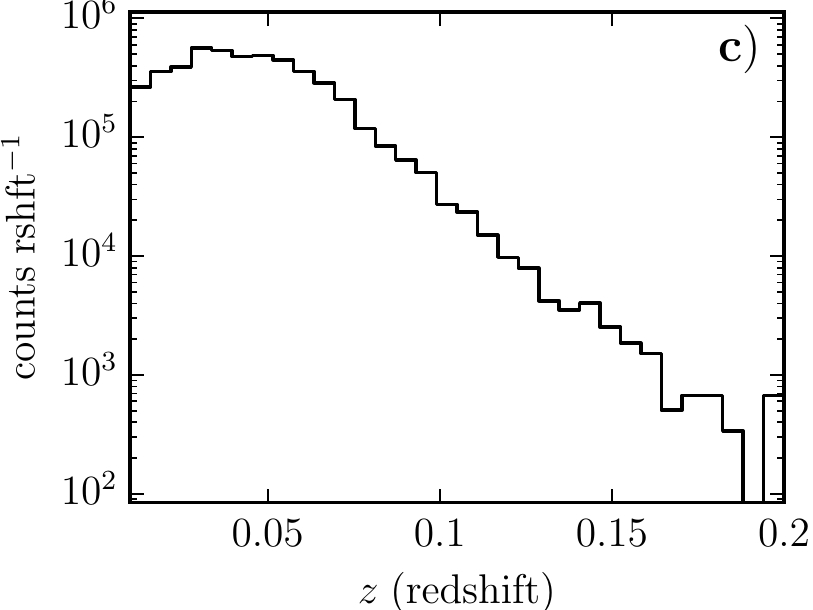}
	
	\caption{Panel \textbf{a} contains a density plot showing the range of luminosities and redshifts sampled by the 6dFGS survey. 
	The dashed lines show luminosity cuts based on the mean SED from \cite{Lake:2016}: any source lower than either line is cut. 
	The blue translucent lines show where the SED variety completeness is 95\% and 98\%, in order of increasing lightness. 
	Panels \textbf{b} and \textbf{c} contain histograms of the data in redshift and luminosity, respectively, after the cut in panel \textbf{a} is applied. 
	No completeness corrections are applied, and the distributions are normalized to the number of data points the set contributes, $N=47,335$ in panel \textbf{a}, and $28,232$ in all other panels.}
	\label{fig:dat:zLuma6d}
\end{figure}

The 6dFGS survey is relatively shallow, as can be seen by the luminosity-redshift graph in Figure~\ref{fig:dat:zLuma6d}, and its marginalized histograms therein, but the coverage is enormous, $1.37\times10^4\operatorname{deg}^2$ after imposing a $\delta < -11.5$ cut to eliminate overlap with SDSS, so the sample size after all limits are imposed is $27,091$.  
Because of this wide coverage, the sample defined here covers galaxies as low as $z=0.01$, but the shallow depth requires an upper limit on the redshifts at $z=0.2$. 
The bright limit imposed on this survey is $K_s > 11.25 \operatorname{AB\ mag}$.

The blue curves in Figure~\ref{fig:dat:zLuma6d} are defined by constant values of the color variety selection function, $S_{\mathrm{color}}(L, z)$ (see Equation~\ref{eqn:dat:colvarcomp}), and bound regions where it is greater than $98\%$ (light blue) and $95\%$ (dark blue). 
They demarcate the regions where the selection function is close enough to constant that the likelihood model defined in LW17I can be neglected. 
For 6dFGS, $95\%$ 17,571 sources ($64.9\%$ of the pre-cut data), and the $98\%$ curve leaves 15,652 ($57.8\%$ of pre-cut). 

\begin{figure}[htb]
	\begin{center}
	\includegraphics[width=0.48\textwidth]{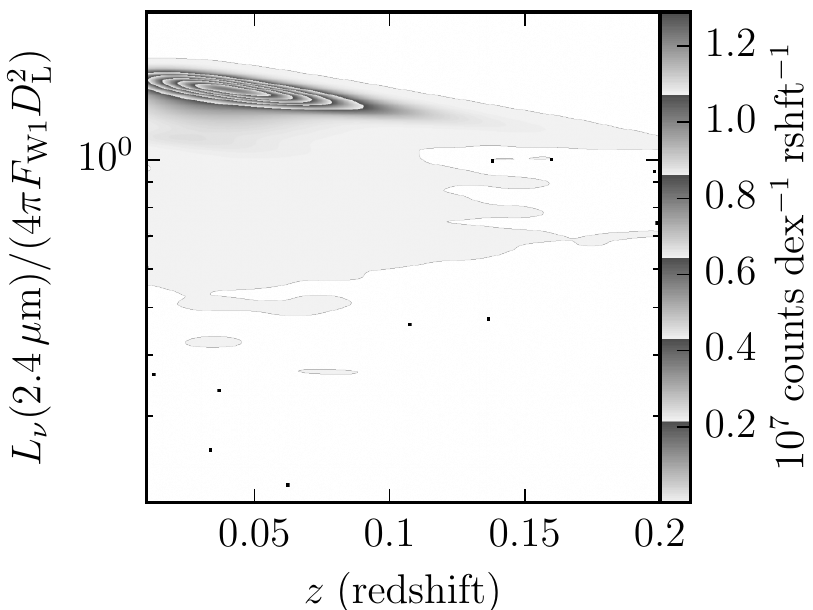}
	\end{center}
	
	\caption{Density plot of K-corrections from W1 to rest frame $2.4\micron$ applied to 6dFGS data.}
	\label{fig:dat:Kcor6d}
\end{figure}

The photometric data used to model galaxy SEDs and define K-corrections, shown in Figure~\ref{fig:dat:Kcor6d}, is summarized in Table~\ref{tbl:dat:photbysurvey}. 
The shape of the data distribution in Figure~\ref{fig:dat:Kcor6d} is consistent with Figure~4 from \cite{Dai:2009}, which used the same set of templates for SED fitting, and Figure~4 from \cite{Blanton:2003LF}. 
The characteristics of the distribution can be explained as the majority of galaxies being fit primarily by the elliptical, Sbc, and irregular templates from Figure~\ref{fig:dat:templates}, that all have nearly the same shape in the region between $1.7$--$3.4\micron$, with a long tail of outliers that are dominated by the AGN template. 

\subsection{SDSS Details}\label{sec:sel:SDSSdets}
The Sloan Digital Sky Survey (SDSS) data release 7, as described in \cite{SDSSdr7}, contained three main extragalactic spectroscopic samples: the main galaxy sample defined in \cite{Strauss:2002}, the red luminous galaxy sample defined in \cite{Eisenstein:2001}, and the quasar sample defined in \cite{Richards:2002}. 
While it would have been nice to be able to use all three samples, the latter two samples are defined using both flux and color cuts, which the model described in LW17I cannot yet accommodate in a timely fashion. 
Only the main galaxy sample is defined in terms of flux ($r \le 17.77 \operatorname{mag}$, Petrosian) and surface brightness ($\mu_{50} \le 24.5\operatorname{mag}\operatorname{arcsec}^{-2}$) in one channel, after extinction correction based on the dust maps from \cite{Schlegel:1998}. 
The sample defined in this paper is therefore limited to the main galaxy sample, with $476,744$ sources after all limits are imposed. 
These limits included cutting out around the survey footprints of the three surveys with significant overlap that were deeper than SDSS, as listed in Table~\ref{tbl:dat:SDSScutouts}. 
There was no overlap with \WD.

Explicitly, in terms of the columns of the SDSS DR10 CasJobs\footnote{\url{http://skyserver.sdss3.org/casjobs/}} database, the selected galaxies had to have: \code{class} set as either ``GALAXY" or ``QSO," $\code{zwarning} = 0$, $(\code{legacy\_target1\,\&\,0x40}) \neq 0$ (that is, the Main Galaxy Sample flag is set, detected using \code{\&} as the `bitwise and' operator), $\code{sdssPrimary} = 1$, and $\code{legacyPrimary} = 1$. 
The exact tables from which we drew data were: \code{SpecObj} for redshifts and flags, \code{SpecDR7} for the magnitudes used for selection, \code{PhotoObj} for additional SDSS photometry, and \code{TwoMassXSC} for 2MASS extended source photometry.

\begin{figure}[htb]
	\includegraphics[width=0.48\textwidth]{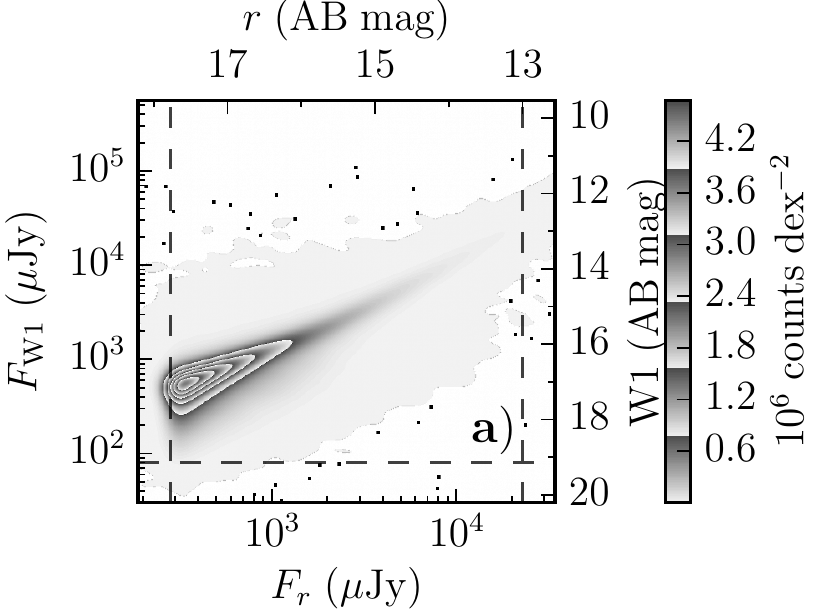}
	\includegraphics[width=0.48\textwidth]{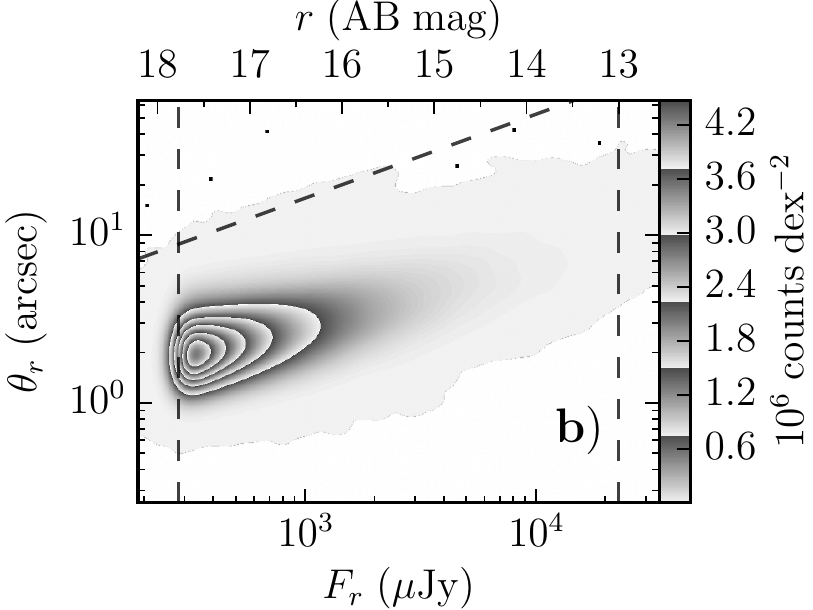}
	
	\caption{Panel \textbf{a} contains a density plot showing the measured fluxes in relationship to the selection limits imposed by the SDSS survey ($17.77 \ge r > 13.0$) and here ($F_{\mathrm{W1}} \ge 80\microJy$). 
	Note how the additional \WISE\ limit eliminates a small fraction of data.
	Panel \textbf{b} contains a density plot showing the measured fluxes in relationship to the selection limits imposed by the SDSS survey ($17.77 \ge r > 10.5$) as the vertical dashed lines and surface brightness within the radius containing half of the Petrosian flux ($0.5 F_r \ge \Sigma_{\mathrm{min}} \pi \theta_r^2$, $-2.5\log_{10}[\Sigma_{\mathrm{min}} / 3631\operatorname{Jy}] = 24.5\operatorname{mag}\operatorname{arcsec}^{-2}$) as the diagonal line in the upper left hand corner of the plot.}
	\label{fig:dat:fluxfluxSD}
\end{figure}

The flux-flux plot is in Figure~\ref{fig:dat:fluxfluxSD}, and it shows that the optical flux limit is the relevant limit for the vast majority of the sources, but not 100\% of them. 
Like 6dFGS, SDSS has a large area on the sky ($7.88\times10^3\operatorname{deg}^2$, after de-overlapping) and so it, too, has a lower redshift limit in this work of $0.01$. 
Likewise, SDSS's shallow depth required upper redshift limit at $z=0.33$, and a bright magnitude limit at $r=13.0\operatorname{mag}$. 
Also in Figure~\ref{fig:dat:fluxfluxSD} is a density plot that shows the relationship of the data to the surface brightness limit, defined as the mean surface brightness within a circle that contains half of the source's Petrosian flux. 
Ideally any analysis would include the surface brightness limit in the selection function model. 
Practically, the surface brightness limit is far from the main body of the data and incorporating it would require an additional measurement of a luminosity-radius relationship that is beyond the scope of the model described in LW17I.

\begin{figure}[htb]
	\includegraphics[width=0.48\textwidth]{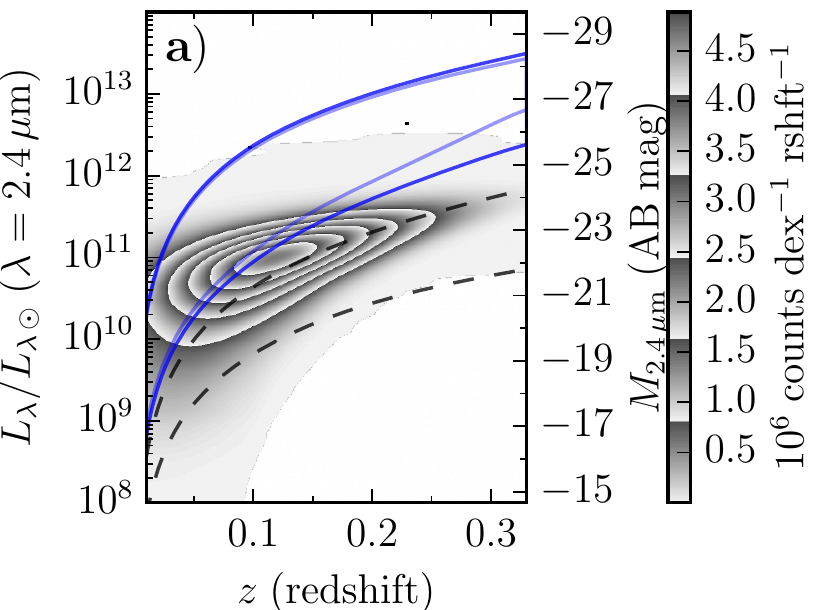}
	\includegraphics[width=0.48\textwidth]{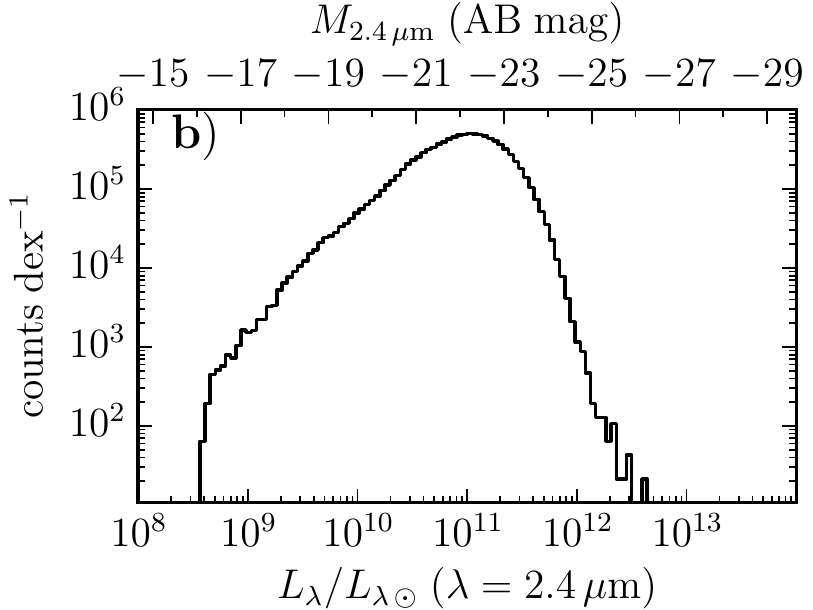}
	\includegraphics[width=0.48\textwidth]{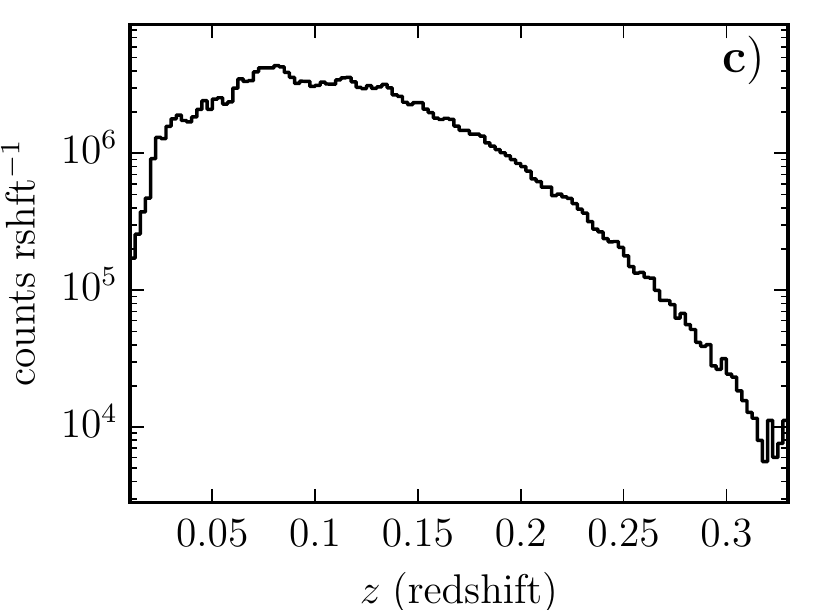}
	
	\caption{Panel \textbf{a} contains a density plot showing the range of luminosities and redshifts sampled by the SDSS survey. 
	The dashed lines show luminosity cuts based on the mean SED from \cite{Lake:2016}: any source lower than either line is cut. 
	The blue translucent lines show where the SED variety completeness is 95\% and 98\%, in order of increasing lightness.
	 Panels \textbf{b} and \textbf{c} contain histograms of the data in redshift and luminosity, respectively, after the cut in panel \textbf{a} is applied.
	 No completeness corrections are applied, and the distributions are normalized to the number of data points the set contributes, $N=626,007$ in panel \textbf{a}, and $476,868$ in all other panels.}
	\label{fig:dat:zLumaSD}
\end{figure}

\begin{figure}[htb]
	\begin{center}
	\includegraphics[width=0.48\textwidth]{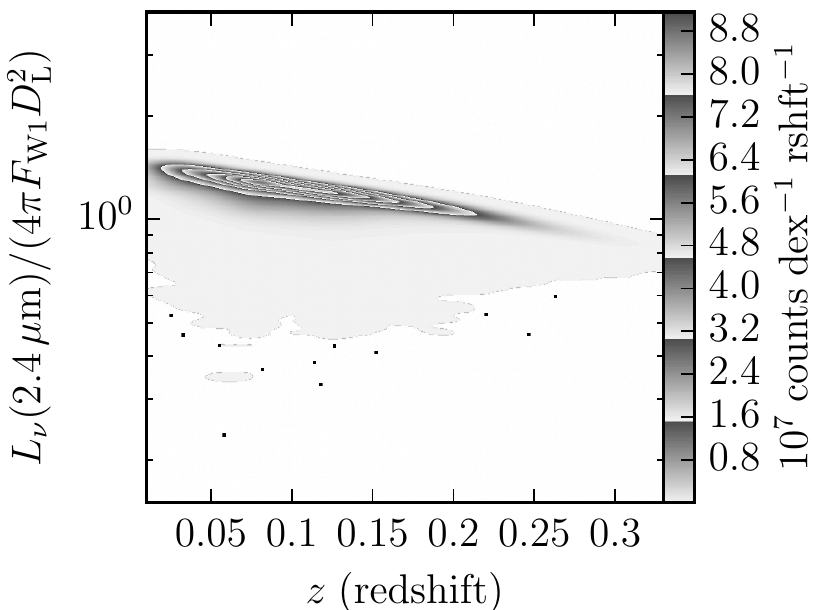}
	\end{center}
	
	\caption{Density plot of K-corrections from W1 to rest frame $2.4\micron$ applied to SDSS data. }
	\label{fig:dat:KcorSD}
\end{figure}

The luminosity-redshift graph is in Figure~\ref{fig:dat:zLumaSD}, along with its marginalizations into histograms. 
The K-corrections applied to calculate those luminosities are shown in Figure~\ref{fig:dat:KcorSD}, and the auxiliary photometric information used to fit the SED models and calculate K-corrections is outlined in Table~\ref{tbl:dat:photbysurvey}. 
The shape of the data distribution in Figure~\ref{fig:dat:KcorSD} is consistent with Figure~4 from \cite{Dai:2009}, which used the same set of templates for SED fitting, and Figure~4 from \cite{Blanton:2003LF}. 
The characteristics of the distribution can be explained as the majority of galaxies being fit primarily by the elliptical, Sbc, and irregular templates from Figure~\ref{fig:dat:templates}, that all have nearly the same shape in the region between $1.7$--$3.4\micron$, with a long tail of outliers that are dominated by the AGN template. 

The blue curves in Figure~\ref{fig:dat:zLumaSD} are defined by constant values of the color variety selection function, $S_{\mathrm{color}}(L, z)$ (see Equation~\ref{eqn:dat:colvarcomp}), and bound regions where it is greater than $98\%$ (light blue) and $95\%$ (dark blue). 
They demarcate the regions where the selection function is close enough to constant that the likelihood model defined in LW17I can be neglected. 
For SDSS, $95\%$ 162,916 sources ($34.2\%$ of the pre-cut data), and the $98\%$ curve leaves 105,835 ($22.2\%$ of pre-cut). 

\floattable
\begin{deluxetable}{lrrrr}
	\tabletypesize{\scriptsize}
	\tablewidth{0.4\textwidth}
	\tablecaption{SDSS Cutouts for Deeper Surveys}
	\tablehead{ \colhead{Survey} & \colhead{$\alpha >$} & \colhead{$\alpha <$} & \colhead{$\delta >$} & \colhead{$\delta <$} \\
		& \colhead{$\deg$} & \colhead{$\deg$} & \colhead{$\deg$} & \colhead{$\deg$} }
	\startdata
		GAMA & $129.00$ & $141.00$ & $-1.00$ & $+3.00$ \\
		GAMA & $174.00$ & $186.00$ & $-2.00$ & $+2.00$ \\
		GAMA & $211.50$ & $223.50$ & $-2.00$ & $+2.00$ \\
		AGES & $216.11$ & $219.77$ & $+32.80$ & $+35.89$ \\
		zCOSMOS & $149.45$ & $150.78$ & $+1.60$ & $+2.86$ \\
	\enddata
	\tablecomments{J2000 right ascension ($\alpha$) and declination ($\delta$) limits for data removed from SDSS for the data described in this work in order to prevent double counting of any sources.}
	\label{tbl:dat:SDSScutouts}
\end{deluxetable}

\subsection{GAMA Details}
The Galaxy And Mass Assembly (GAMA) survey was originally defined over three fields for the first data release, described in \cite{Baldry:2010}, and later expanded to 5 fields, as described in the second data release paper, \cite{Liske:2015}. 
After the final data release the depth at which the data is complete will vary depending on the field. 
For the second data release, all of the fields are complete down to at least $r = 19\operatorname{mag}$, extinction corrected Petrosian, and so that is the limit used for the selection in this paper. 
Selecting science quality redshifts from GAMA is relatively straightforward, as the GAMA team supplies a `normalized quality' integer, \code{NQ}. 
The high quality redshifts satisfy: $\code{NQ} > 2$.

\begin{figure}[htb]
	\includegraphics[width=0.48\textwidth]{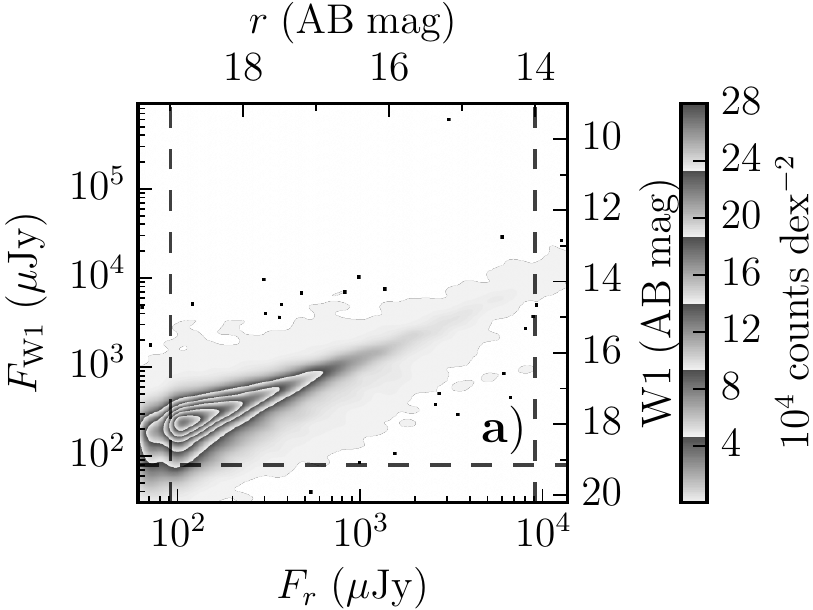}
	\includegraphics[width=0.48\textwidth]{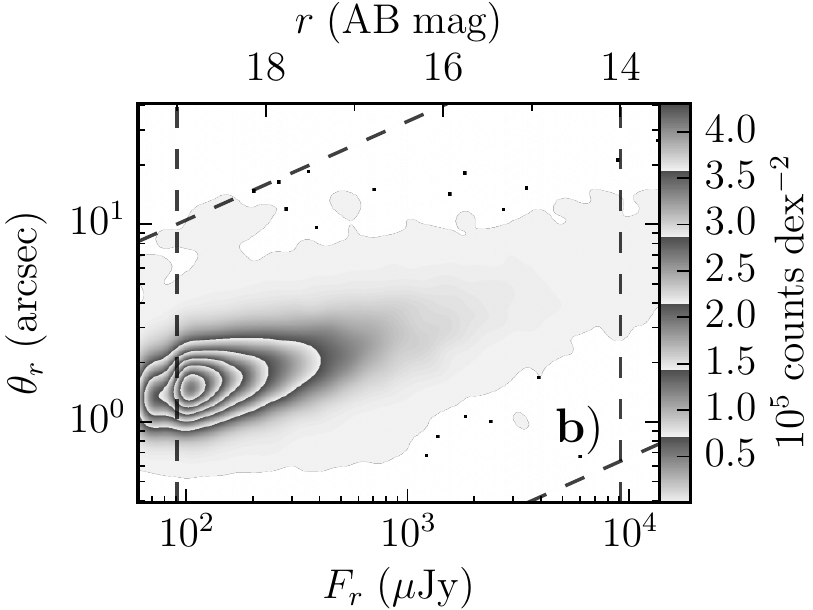}
	
	\caption{Panel \textbf{a} contains a density plot showing the measured fluxes in relationship to the selection limits imposed by the GAMA survey ($19 \ge r > 10$) and here ($F_{\mathrm{W1}} \ge 80\microJy$). 
	Note how the additional \WISE\ limit eliminates a noticeable fraction of the data. 
	Panel \textbf{b} contains a density plot showing the measured fluxes in relationship to the selection limits imposed by the GAMA survey ($19 \ge r > 14$) as the vertical dashed lines and surface brightness within the radius containing half of the Petrosian flux ($\Sigma_{\mathrm{max}} \pi \theta_r^2 \ge 0.5 F_r \ge \Sigma_{\mathrm{min}} \pi \theta_r^2 $, $-2.5\log_{10}[\Sigma_{\mathrm{min}} / 3631\operatorname{Jy}] = 24.5\operatorname{mag}\operatorname{arcsec}^{-2}$, $-2.5\log_{10}[\Sigma_{\mathrm{max}} / 3631\operatorname{Jy}] = 15\operatorname{mag}\operatorname{arcsec}^{-2}$) as the diagonal line in the upper left hand corner of the plot.}
	\label{fig:dat:fluxfluxGA}
\end{figure}

As can be seen in Figure~\ref{fig:dat:fluxfluxGA}, the optical limit is the controlling one for the vast majority of sources, but the W1 flux limit is relevant for a sizable fraction of the galaxies in GAMA. 
Like SDSS, GAMA imposes surface brightness limits, both high and low. 
Their relationship to the data can be found in Figure~\ref{fig:dat:fluxfluxGA}, and just like for the SDSS subsample, it is beyond the scope of the model described in LW17I to account for these limits. 
After all limits are imposed, this survey contributes 44,495 sources to the sample.

\begin{figure}[htb]
	\includegraphics[width=0.48\textwidth]{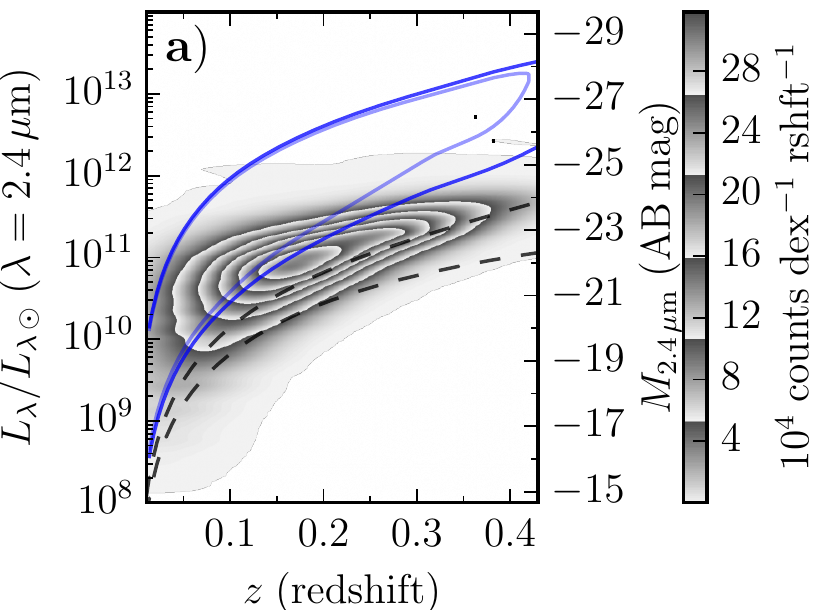}
	\includegraphics[width=0.48\textwidth]{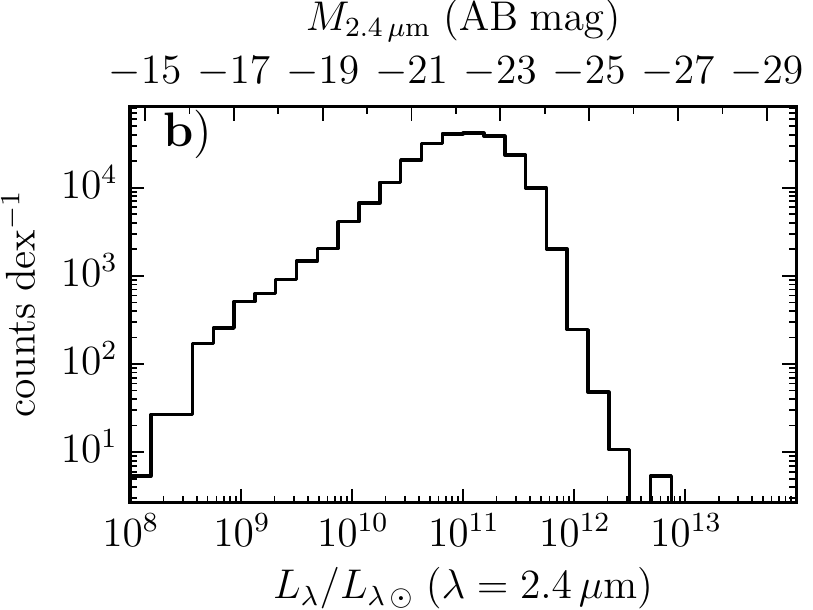}
	\includegraphics[width=0.48\textwidth]{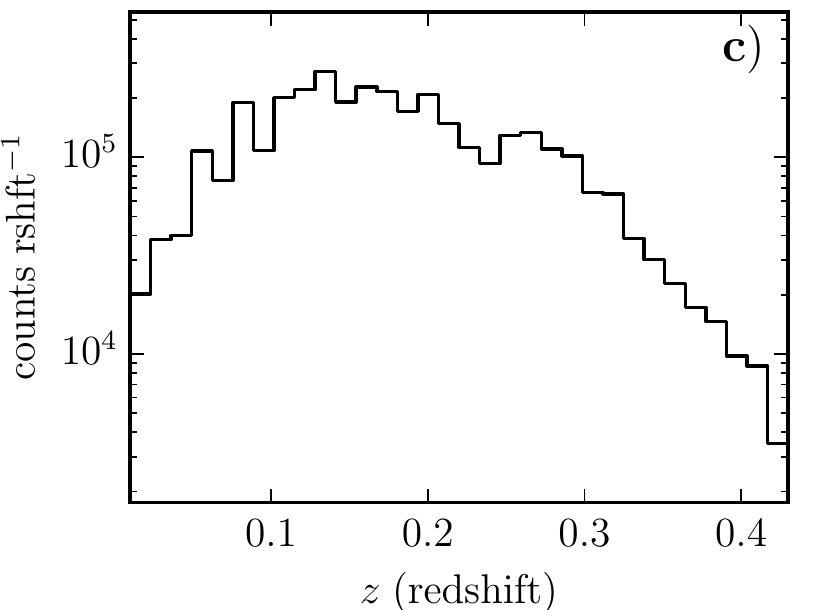}
	
	\caption{Panel \textbf{a} contains a density plot showing the range of luminosities and redshifts sampled by the GAMA survey. 
	The dashed lines show luminosity cuts based on the mean SED from \cite{Lake:2016}: any source lower than either line is cut. 
	Panels \textbf{b} and \textbf{c} contain histograms of the data in redshift and luminosity, respectively, after the cut in panel \textbf{a} is applied. 
	The blue translucent lines show where the SED variety completeness is 95\% and 98\%, in order of increasing lightness. 
	No completeness corrections are applied, and the distributions are normalized to the number of data points the set contributes, $N=52,773$ in panel \textbf{a}, and $44,604$ in all other panels.}
	\label{fig:dat:zLumaGA}
\end{figure}

The luminosity versus redshift density plot, found in Figure~\ref{fig:dat:zLumaGA} with its marginalizations, shows that GAMA is the shallowest survey to significantly sample galaxies from the median redshift of the \WD\ survey. 
It is also the narrowest survey for which the selection defined in this paper covers redshifts down to $z=0.01$, and for which a low maximum redshift was imposed at $z=0.43$. 
The bright limit imposed here was at $r=14\operatorname{mag}$.

The blue curves in Figure~\ref{fig:dat:zLumaGA} are defined by constant values of the color variety selection function, $S_{\mathrm{color}}(L, z)$ (see Equation~\ref{eqn:dat:colvarcomp}), and bound regions where it is greater than $98\%$ (light blue) and $95\%$ (dark blue). 
They demarcate the regions where the selection function is close enough to constant that the likelihood model defined in LW17I can be neglected. 
For GAMA, $95\%$ 15,659 sources ($35.2\%$ of the pre-cut data), and the $98\%$ curve leaves 9,641 ($21.7\%$ of pre-cut). 

\begin{figure}[htb]
	\begin{center}
	\includegraphics[width=0.48\textwidth]{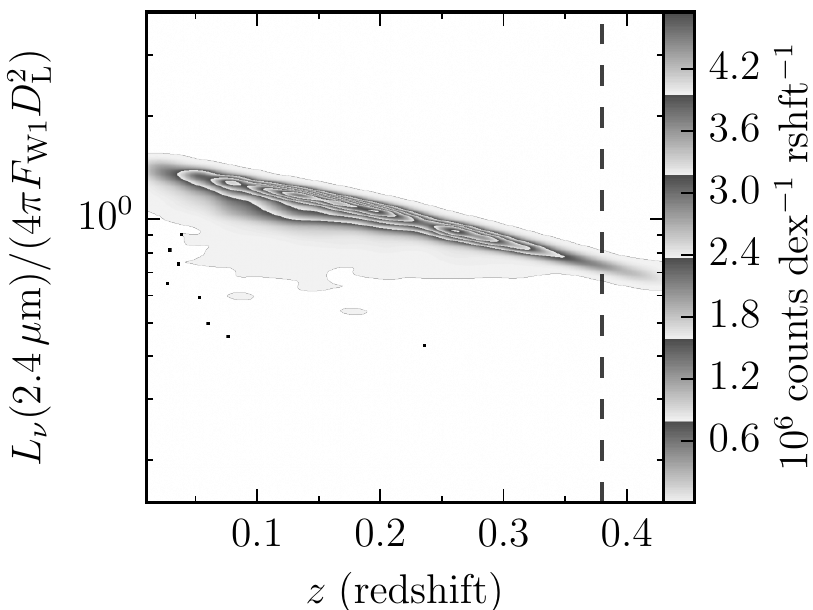}
	\end{center}
	
	\caption{Density plot of K-corrections from W1 to rest frame $2.4\micron$ applied to GAMA data. 
	The vertical dashed line at $z=0.38$ marks the fiducial redshift used in this work, where the K-correction is only the bandwidth scaling by $1+z$.}
	\label{fig:dat:KcorGA}
\end{figure}

The K-corrections applied to calculate the luminosities are shown in Figure~\ref{fig:dat:KcorGA}, and the photometry used to fit the SEDs used to calculate the K-corrections are summarized in Table~\ref{tbl:dat:photbysurvey}. 
The shape of the data distribution in Figure~\ref{fig:dat:KcorGA} is consistent with Figure~4 from \cite{Dai:2009}, which used the same set of templates for SED fitting, and Figure~4 from \cite{Blanton:2003LF}. 
The characteristics of the distribution can be explained as the majority of galaxies being fit primarily by the elliptical, Sbc, and irregular templates from Figure~\ref{fig:dat:templates}, that all have nearly the same shape in the region between $1.7$--$3.4\micron$, with a long tail of outliers that are dominated by the AGN template. 

\subsection{AGES Details}
The AGN and Galaxy Evolution Survey (AGES), described in \cite{AGES}, is a spectroscopic redshift survey targeted using photometry from the NOAO Deep, Wide-Field, Survey (NDWFS) and the \textit{Spitzer} Deep, Wide-Field, Survey (SDWFS). 
\cite{AGES} defines a lot of subsamples with different flux-limits. 
The sample defined in this paper uses the main $I$-band selected sample, defined with $(\code{code06 \& 0x80000}) \neq 0$ (\code{0x80000} is a hexadecimal integer equal to $2^{20}$ in base $10$) in \cite{AGES}, which is defined by $I$-band flux limits to be complete brighter than $I=18.9\operatorname{mag}$, and 20\% complete below that down to $I=20.4\operatorname{mag}$. 
The AGES data with well analyzed completeness is limited to a set of 15 overlapping circular fields, but the released redshifts cover a larger area. 
Limiting the sample to just those redshifts in the canonical fields requires selecting sources with $\code{field} > 0$.

\begin{figure}[htb]
	\begin{center}
	\includegraphics[width=0.48\textwidth]{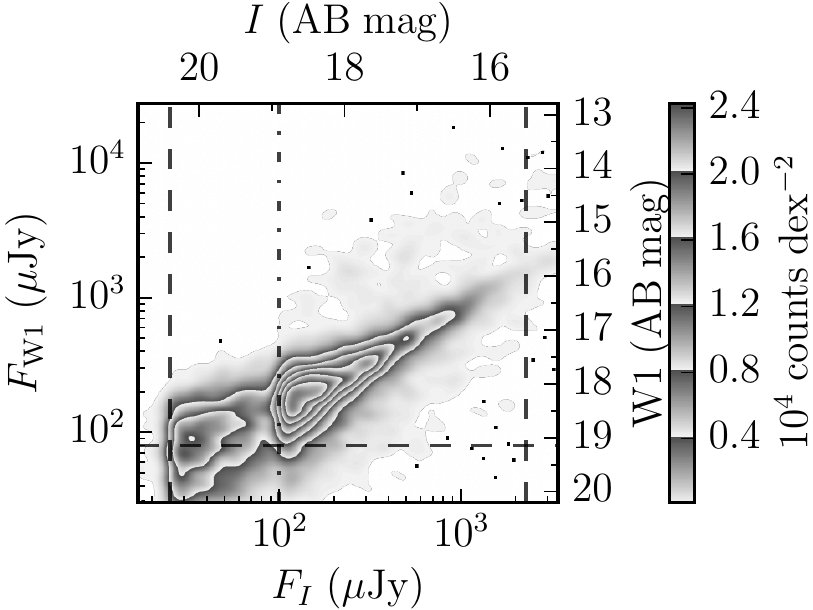}
	\end{center}
	
	\caption{Density plot showing the measured fluxes in relationship to the selection limits imposed by the AGES survey ($20.4 \ge I > 15.5$ with 20\% completeness setting in fainter than $I = 18.9$) and here ($F_{\mathrm{W1}} \ge 80\microJy$). 
	Note how this is the first survey for which the main locus on which galaxies lie intersects the \WISE\ limit imposed here.}
	\label{fig:dat:fluxfluxAG}
\end{figure}

The relationship of the data to the flux limits are shown in Figure~\ref{fig:dat:fluxfluxAG}. 
This is the survey for which taking into account both the optical and W1 flux limits is most important because the locus on which most galaxies are found goes into the corner defined by the flux limits. 
AGES is narrow enough that the sample defined this paper only includes data with redshifts $z > 0.05$ and deep enough for redshifts out to $z=1$. 
The bright limit we imposed is at $I=15.5\operatorname{mag}$. 
After all limits are imposed, AGES contributes 6,588 galaxies to the sample defined in this paper.

The NDWFS astrometry has a known astrometric offset relative to other major surveys. 
So before performing the final $3\arcsec$ distance match cut, we calculated the mean offset of nearest neighbors and used that offset to correct the distance between the AllWISE source and the NDWFS sources. 
The offsets we found were:$-0.29\arcsec$ in right ascension, and $-0.14\arcsec$ in declination.

\begin{figure}[htb]
	\includegraphics[width=0.48\textwidth]{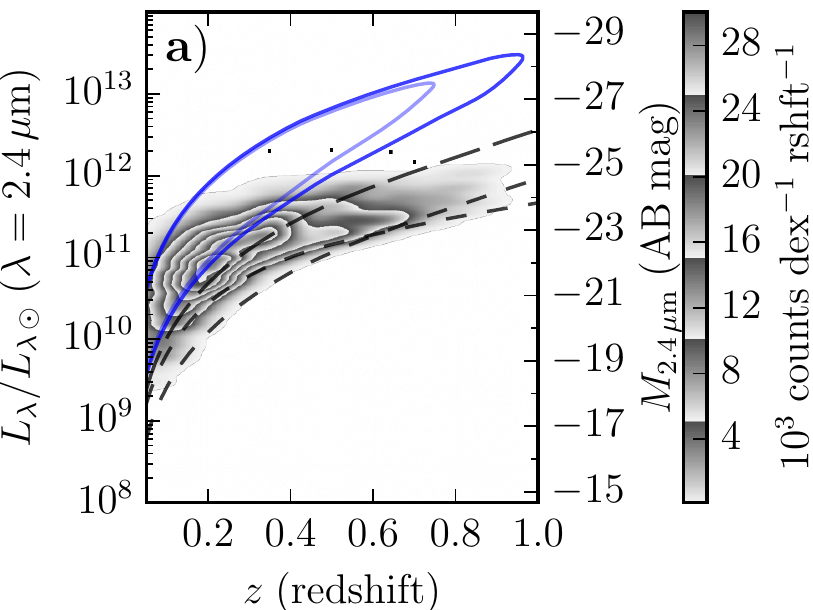}
	\includegraphics[width=0.48\textwidth]{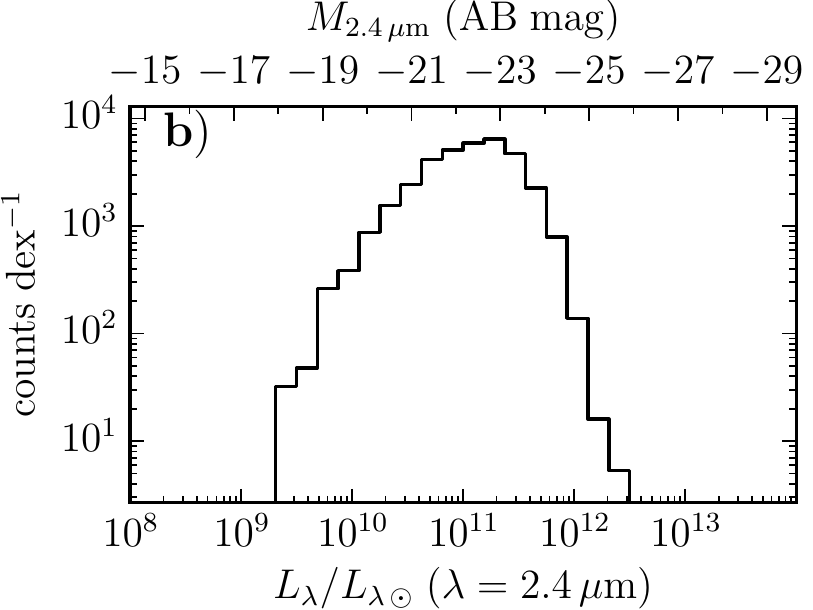}
	\includegraphics[width=0.48\textwidth]{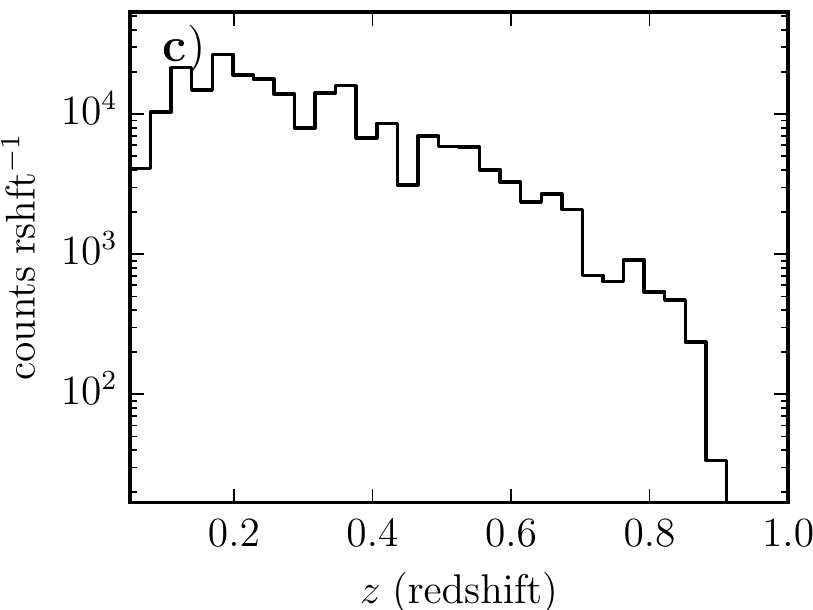}
	
	\caption{Panel \textbf{a} contains a density plot showing the range of luminosities and redshifts sampled by the AGES survey. 
	The short dashed lines show luminosity cuts based on the mean SED from \cite{Lake:2016}: any source lower than either short dashed line is cut, and any source brighter than $I=18.9 \operatorname{AB\,mag}$ and below the long dashed line is cut. 
	The blue translucent lines show where the SED variety completeness is 95\% and 98\%, in order of increasing lightness. 
	Panels \textbf{b} and \textbf{c} contain histograms of the data in redshift and luminosity, respectively, after the cut in panel \textbf{a} is applied. 
	No completeness corrections are applied, and the distributions are normalized to the number of data points the set contributes, $N=6,603$ in panel \textbf{a}, and $6,553$ in all other panels.}
	\label{fig:dat:zLumaAG}
\end{figure}

The density plot showing the luminosities versus redshift is in Figure~\ref{fig:dat:zLumaAG}, alongside its marginalizations. 
The blue curves in Figure~\ref{fig:dat:zLumaAG} are defined by constant values of the color variety selection function, $S_{\mathrm{color}}(L, z)$ (see Equation~\ref{eqn:dat:colvarcomp}), and bound regions where it is greater than $98\%$ (light blue) and $95\%$ (dark blue). 
The faint sample is automatically excluded from these regions because it is too narrow to provide a flat selection region. 
They demarcate the regions where the selection function is close enough to constant that the likelihood model defined in LW17I can be neglected. 
For AGES, $95\%$ 2,096 sources ($36.5\%$ of the pre-cut data), and the $98\%$ curve leaves 1,664 ($29.0\%$ of pre-cut). 

\begin{figure}[htb]
	\begin{center}
	\includegraphics[width=0.48\textwidth]{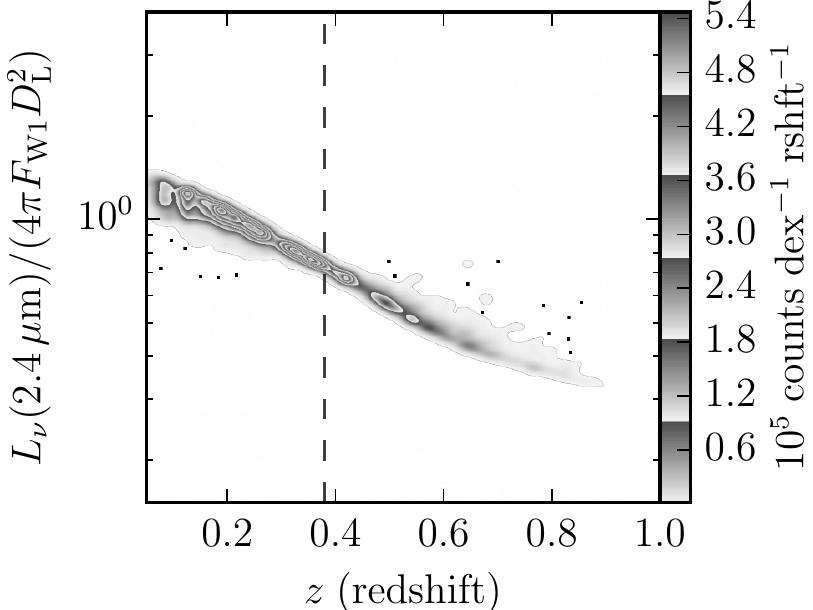}
	\end{center}
	
	\caption{Density plot of K-corrections from W1 to rest frame $2.4\micron$ applied to AGES data. 
	The vertical dashed line at $z=0.38$ marks the fiducial redshift used in this work, where the K-correction is only the bandwidth scaling by $1+z$.}
	\label{fig:dat:KcorAG}
\end{figure}

The K-corrections used to calculate luminosities for AGES galaxies are shown in Figure~\ref{fig:dat:KcorAG}, and the photometric data used to fit the SEDs for calculating the K-corrections is summarized in Table~\ref{tbl:dat:photbysurvey}. 
The shape of the data distribution in Figure~\ref{fig:dat:KcorAG} is consistent with Figure~4 from \cite{Dai:2009}, which used the same set of templates for SED fitting, and Figure~4 from \cite{Blanton:2003LF}. 
The characteristics of the distribution can be explained as the majority of galaxies being fit primarily by the elliptical, Sbc, and irregular templates from Figure~\ref{fig:dat:templates}, that all have nearly the same shape in the region between $1.7$--$3.4\micron$, with a long tail of outliers that are dominated by the AGN template. 

The photometry published with the main AGES paper, \cite{AGES}, did not include uncertainties, so we performed a cross-match against NDWFS and SDWFS (the combined epoch IRAC c1 driven extraction stack only). 
The AGES sources didn't always have a counterpart in the NDWFS and SDWFS catalogs. 
In the case of SDWFS, that is because the data release used here is newer than the one used for AGES and this work only used the c1 stack catalog. 
Noise models were, therefore, also fit to the data to produce model uncertainties when a catalog uncertainty was unavailable. 
The noise model takes the form of a smoothly broken power law:
\begin{align}
	\sigma(F) & = \sigma_{\mathrm{knee}} \left(\frac{F}{F_{\mathrm{knee}}}\right)^\alpha \left(\frac{1}{2} + \frac{1}{2}\cdot \left[\frac{F}{F_{\mathrm{knee}}}\right]^{|\beta|s}\right)^{\operatorname{sign}(\beta) / s},\label{eqn:SBPLnoise}
\end{align}
where $F_{\mathrm{knee}}$ is the location of the break, or knee, in the power law, $\sigma_{\mathrm{knee}} \equiv \sigma(F_\mathrm{knee})$, $\alpha$ is the faint end slope, $\beta$ is the change in slope at $F_{\mathrm{knee}}$, and $s$ is a positive parameter setting the sharpness of the break. 
For $s\rightarrow 0$ the break becomes infinitely wide, and for $s\rightarrow \infty$ it becomes infinitely sharp (that is, a corner). 
The noise model parameters found from fitting individual bands, after trimming outliers, are listed in Table~\ref{tbl:datAGnoiseMod}.

\begin{deluxetable}{cccccc}
	\tabletypesize{\scriptsize}
	\tablewidth{0.4\textwidth}
	\tablecaption{AGES Error Models}
	\tablehead{\colhead{Band} & \colhead{$F_{\mathrm{knee}}$} & \colhead{$\sigma_{\mathrm{knee}}$} & \colhead{$\alpha$} & \colhead{$\beta$} & \colhead{$s$} \\
	& \colhead{$\microJy$} & \colhead{$\microJy$} & & & }
	\startdata
	$B_w$ & $0.13$& $0.017$ & 0.51 & -0.39 & 6.5 \\
	$R$ & $0.76$& $0.010$ & 0.58 & -0.47 & 8.3 \\
	$I$ & $1.3$& $0.15$ & 0.49 & -0.38 & 14 \\
	$K$ & $48$& $8.3$ & 0.80 & -0.56 & 5.9 \\
	c1 & $5.3$ & $0.99$ & 0.67 & -0.59 & 6.7 \\
	c2\tablenotemark{a}  & $1$& $1.3$ & 0.10 & $0$ & $0$ \\
	c3\tablenotemark{a}  & $1$ & $7.3$ & 0.042 & $0$ & $0$ \\
	c4\tablenotemark{a}  & $1$ & $7.8$ & 0.025 & $0$ & $0$ \\
	\enddata
	\tablecomments{Noise model parameters used to compute flux uncertainties in the absence of uncertainties from NDWFS or SDWFS, as defined in Equation~\ref{eqn:SBPLnoise}.}
	\label{tbl:datAGnoiseMod}
	\tablenotetext{a}{The data for this channel did not exhibit a knee, so a power law fit was used instead.}
\end{deluxetable}

\subsection{zCOSMOS Details}\label{sec:sel:zcos}
The sample defined in this paper uses the subset of the zCOSMOS survey known as the ``10k-Bright Spectroscopic Sample," is described in \cite{Lilly:2009} and \cite{Knobel:2012}. 
The COSMOS field has been the subject of an intensive campaign of imaging by many groups, as described in \cite{Scoville:2007}. 
zCOSMOS based its targeting on photometry from Hubble Advanced Camera for Surveys (ACS) Wide Field Camera (WFC) imaging with the F814W filter, which is approximately $I$-band. 
The 10k, data release 2, subset of the survey is 62\% complete for compulsory targets, and 30\% complete for the rest. 
Selecting high quality redshifts from zCOSMOS is the most involved of the surveys used here because of the detailed `confidence class' (\code{cc}) system used. 
The recommendation in \cite{Lilly:2009} is to accept all sources with \code{cc} equal to: any 3.X, 4.X, 1.5, 2.4, 2.5, 9.3, and 9.5. 
Based on the description of those classes, the sample defined here accepts sources that fit in the recommended classes, but also those with a leading 1 (10 was added to show broad line AGN), 18.3, 18.5, and to reject all secondary targets (2 in the tens or hundreds digit). 
This can be done by accepting sources for which the text string version of \code{cc} matches the regular expression ``\verb=([34]\..*)|([1289]\.5)|(2\.4)|([89]\.3)=" and doesn't match ``\verb=^2\d+\.=". 
Finally, the targets fell into three selection classes, column named \code{i}, and `unintended' sources are rejected by requiring $\code{i}>0$.

\begin{figure}[htb]
	\begin{center}
	\includegraphics[width=0.48\textwidth]{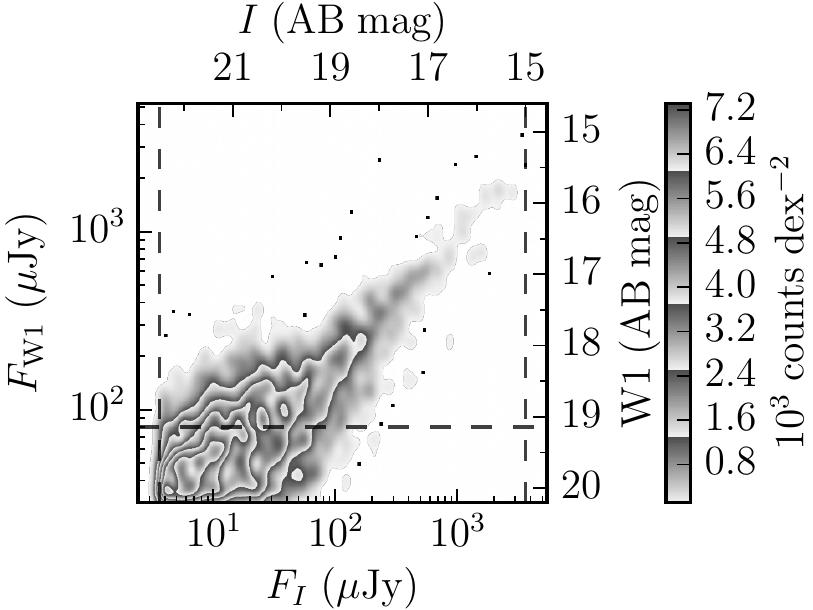}
	\end{center}
	
	\caption{Density plot showing the measured fluxes in relationship to the selection limits imposed by the zCOSMOS survey ($22.5 \ge I > 15.0$) and here ($F_{\mathrm{W1}} \ge 80\microJy$). 
	Note how the majority of the data is eliminated by the \WISE\ limit we imposed, though the data is still near the intersection of the limits.}
	\label{fig:dat:fluxfluxzC}
\end{figure}

As Figure~\ref{fig:dat:fluxfluxzC} shows, even zCOSMOS is affected by the need to use both W1 and $I$-band limits in the analysis of the data. 
Like AGES, the narrowness of zCOSMOS means that the sample herein is limited to redshifts $z > 0.05$. 
After all limits are imposed, this survey contributes 1,267 galaxies to the sample.

\begin{figure}[htb]
	\includegraphics[width=0.48\textwidth]{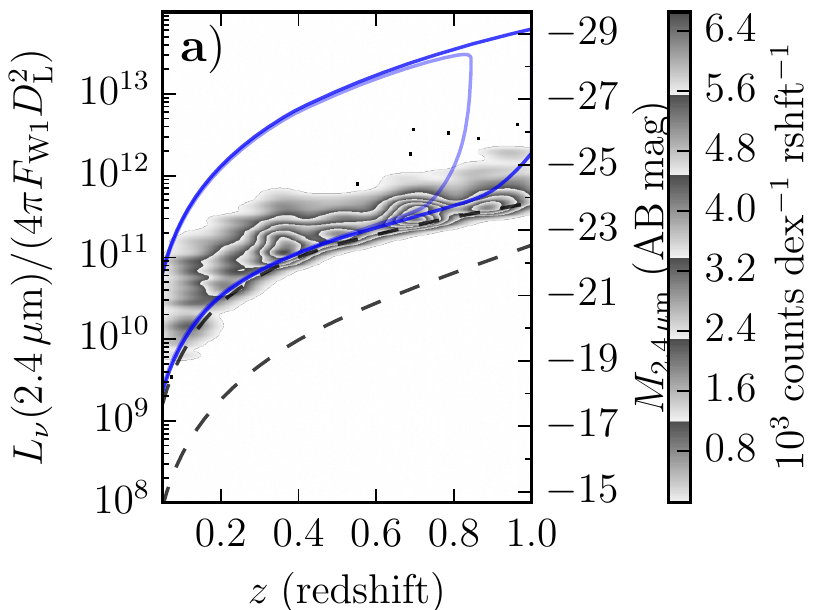}
	\includegraphics[width=0.48\textwidth]{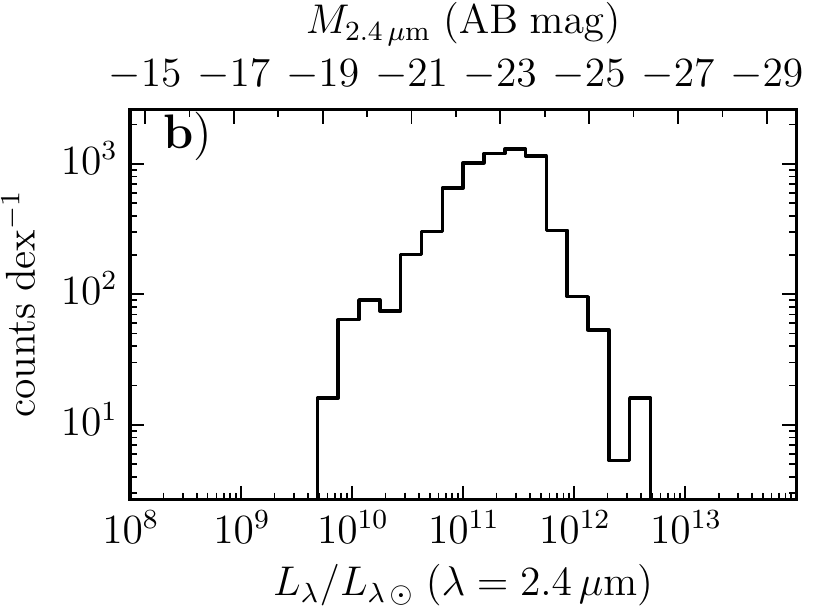}
	\includegraphics[width=0.48\textwidth]{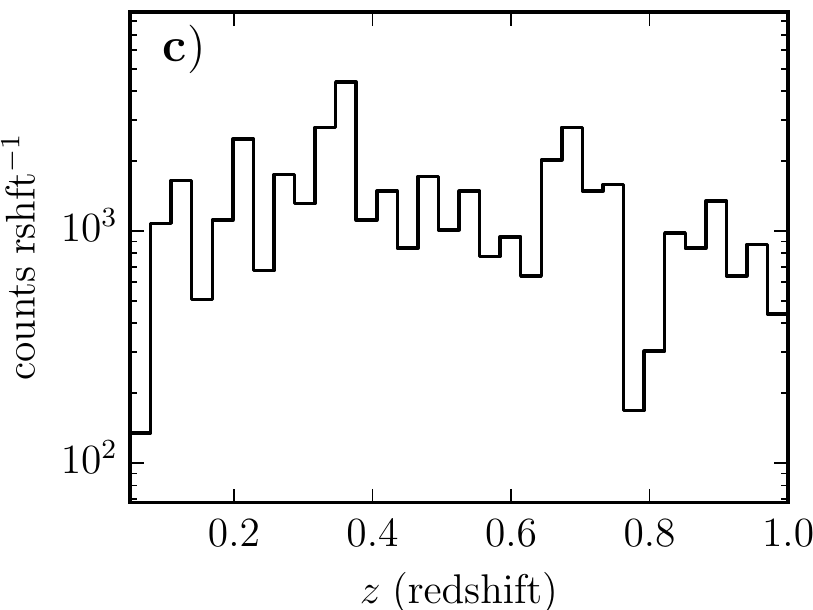}
	
	\caption{Panel \textbf{a} contains a density plot showing the range of luminosities and redshifts sampled by the zCOSMOS survey. 
	The dashed lines show luminosity cuts based on the mean SED from \cite{Lake:2016}: any source lower than either line is cut. 
	The blue translucent lines show where the SED variety completeness is 95\% and 98\%, in order of increasing lightness.
	Panels \textbf{b} and \textbf{c} contain histograms of the data in redshift and luminosity, respectively, after the cut in panel \textbf{a} is applied.
	No completeness corrections are applied, and the distributions are normalized to the number of data points the set contributes, $N=1,301$ in panel \textbf{a}, and $1,231$ in all other panels.}
	\label{fig:dat:zLumazC}
\end{figure}

The density plot showing the data in luminosity-redshift space is in Figure~\ref{fig:dat:zLumazC}, alongside its marginalizations. 
The blue curves in Figure~\ref{fig:dat:zLumazC} are defined by constant values of the color variety selection function, $S_{\mathrm{color}}(L, z)$ (see Equation~\ref{eqn:dat:colvarcomp}), and bound regions where it is greater than $98\%$ (light blue) and $95\%$ (dark blue). 
They demarcate the regions where the selection function is close enough to constant that the likelihood model defined in LW17I can be neglected. 
For zCOSMOS, $95\%$ 890 sources ($72.7\%$ of the pre-cut data), and the $98\%$ curve leaves 763 ($62.3\%$ of pre-cut). 

\begin{figure}[htb]
	\begin{center}
	\includegraphics[width=0.48\textwidth]{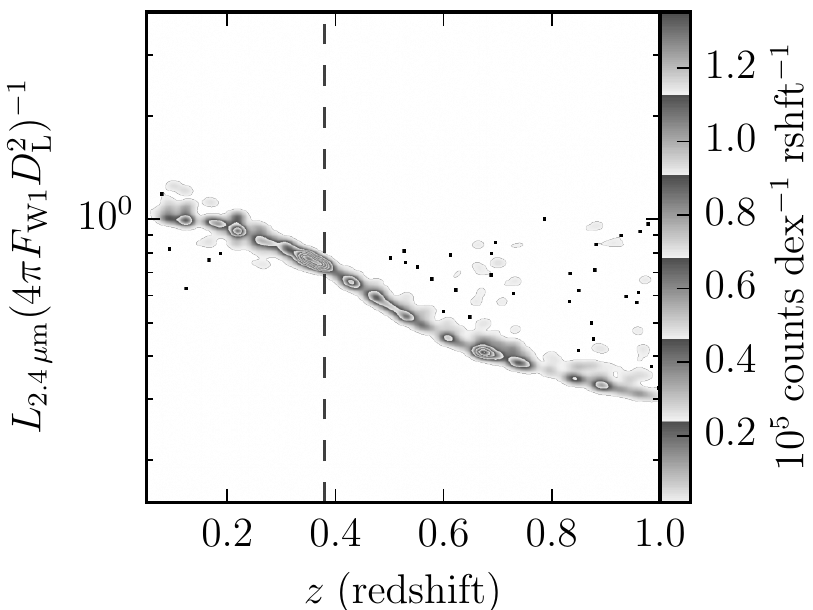}
	\end{center}
	
	\caption{Density plot of K-corrections applied from W1 to rest frame $2.4\micron$ to zCOSMOS data. 
	The vertical dashed line at $z=0.38$ marks the fiducial redshift used in this work, where the K-correction is only the bandwidth scaling by $1+z$.}
	\label{fig:dat:KcorzC}
\end{figure}

The K-corrections used to calculate those luminosities are shown in Figure~\ref{fig:dat:KcorzC}, and the photometric information used to fit the SEDs used to calculate the K-corrections are summarized in Table~\ref{tbl:dat:photbysurvey}. 
The shape of the data distribution in Figure~\ref{fig:dat:KcorzC} is consistent with Figure~4 from \cite{Dai:2009}, which used the same set of templates for SED fitting, and Figure~4 from \cite{Blanton:2003LF}. 
The characteristics of the distribution can be explained as the majority of galaxies being fit primarily by the elliptical, Sbc, and irregular templates from Figure~\ref{fig:dat:templates}, that all have nearly the same shape in the region between $1.7$--$3.4\micron$, with a long tail of outliers that are dominated by the AGN template. 

\section{Post Selection Data Tables} \label{sec:dat}
Each cross-matched survey has its own layout, but the general layout is as follows: target coordinates in decimal degrees (J2000 right ascension and declination), the unique object identifier from the redshift survey (if provided), the identifiers from the photometric surveys to which the object successfully matched, the parameters from the template fits (including the $\chi^2$ of the fit and the formal number of degrees of freedom; ignoring the impact that the constraints on the parameters have on that number), and the boundary redshifts for inclusion in this data set (including an intermediate redshift, \code{z\_mid}, for sources in \WD\ or AGES to account for the intermediate flux cuts of those surveys). 
In order to keep file size down, that is the extent of the information published with this work. 
The rest of the information, like the redshift and photometric properties, is available from the various original sources referenced in the tables of Section~\ref{sec:sel}. 
Excerpts from the machine readable tables can be found in: Tables~\ref{tbl:dat:modWD1} and \ref{tbl:dat:modWD2} for \WD, Tables~\ref{tbl:dat:mod6d1} and \ref{tbl:dat:mod6d2} for 6dFGS, Tables~\ref{tbl:dat:modSD1} and \ref{tbl:dat:modSD2} for SDSS, Tables~\ref{tbl:dat:modGA1} and \ref{tbl:dat:modGA2} for GAMA, Tables~\ref{tbl:dat:modAG1} and \ref{tbl:dat:modAG2} for AGES, and Tables~\ref{tbl:dat:modzC1} and \ref{tbl:dat:modzC2} for zCOSMOS.

\floattable
\begin{deluxetable}{rrrrrr}
	\tabletypesize{\scriptsize}
	\tablewidth{0.9\textwidth}
	\tablecaption{Excerpt from \WD\ Extended Data, Column Set 1}
	\tablehead{ \colhead{\code{Ra}} & \colhead{\code{Dec}} & \colhead{\code{WDID}} & 
		\colhead{\code{GALEX\_ID}} & \colhead{\code{SDSS\_ID}} & \colhead{\code{AllWISE\_ID}}  \\
		\colhead{$\deg$} & \colhead{$\deg$} &  &
		&  &  
		}
	\startdata
		$310.319670$ & $-14.525610$ & 19 & null & 1237668758314746796 & 3099m152\_ac51-053725 \\
		$310.367640$ & $-14.514580$ & 23 & null & 1237668758314746981 & 3099m152\_ac51-053735 \\
		$310.315280$ & $-14.509710$ & 27 & null & 1237668758314747310 & 3099m152\_ac51-053770 \\
		$310.217560$ & $-14.419560$ & 59 & null & 1237668758314681889 & 3099m152\_ac51-053534 \\
		$312.385500$ & $-11.708410$ & 102 & 6379641521644244188 & null & 3121m122\_ac51-047591 \\
		$312.394260$ & $-11.672530$ & 153 & null & null & 3121m122\_ac51-048520 \\
		$312.344700$ & $-11.667790$ & 155 & null & null & 3121m122\_ac51-047825 \\
	\enddata
	\tablecomments{ First set of \WD\ extended data columns. 
	The first two columns are the right ascension and declination, in J2000 decimal degrees. 
	The \code{WDID} column is an integer index uniquely assigned to targets in the \WD\ survey. 
	\code{GALEX\_ID} is a uniquely identifying integer assigned in GALEX gr7 to the source (\code{null} if not matched). \code{SDSS\_ID} is an integer assigned to the matched source in SDSS data release 10 (\code{null} if not matched). 
	\code{AllWISE\_ID} is the \href{http://wise2.ipac.caltech.edu/docs/release/allwise/expsup/sec2\_1a.html\#source\_id}{\code{source\_id}} assigned to the source in the AllWISE survey.  }
	\label{tbl:dat:modWD1}
\end{deluxetable}

\floattable
\begin{deluxetable}{r|rrrrrrrrrrr}
	\tabletypesize{\scriptsize}
	\rotate
	\tablewidth{0.8\textheight}
	\tablecaption{Excerpt from \WD\ Extended Data, Column Set 2}
	\tablehead{ \colhead{\code{WDID}} & 
		\colhead{\code{Ell}} & \colhead{\code{Sbc}} & \colhead{\code{Irr}} & \colhead{\code{AGN}} & 
		\colhead{\code{AGN\_EBmV}} & \colhead{\code{chisqr}} & \colhead{\code{Ndf}} & 
		\colhead{\code{FitMode}} & 
		\colhead{\code{z\_min}} & \colhead{\code{z\_mid}} & \colhead{\code{z\_max}} \\
		 & 
		\colhead{$10^{10}\, L_\odot$} & \colhead{$10^{10}\, L_\odot$} & \colhead{$10^{10}\, L_\odot$} & \colhead{$10^{10}\, L_\odot$} &
		\colhead{$\operatorname{mag}$} &  &  &
		&
		&  &
		}
	\startdata
		19 & $3.4370$e$-02$ & $0.0000$e$+00$ & $9.9800$e$-01$ & $1.0930$e$-03$ & $0.1865$ & $5.100$e$-27$ & -2 & main & $0.029$ & $0.114$ & $0.138$ \\
		23 & $7.6680$e$+00$ & $0.0000$e$+00$ & $9.7870$e$-01$ & $7.9540$e$-01$ & $0.0000$ & $1.500$e$+01$ & 3 & main & $0.060$ & $0.472$ & $0.579$ \\
		27 & $0.0000$e$+00$ & $4.2510$e$+01$ & $3.5730$e$+00$ & $0.0000$e$+00$ & $0.0000$ & $7.000$e$+01$ & 1 & main & $0.090$ & $0.653$ & $0.819$ \\
		29 & $0.0000$e$+00$ & $4.8780$e$+00$ & $6.8260$e$+00$ & $1.2160$e$-01$ & $10.5400$ & $2.700$e$-29$ & -2 & main & $0.077$ & $0.353$ & $0.425$ \\
		34 & $0.0000$e$+00$ & $0.0000$e$+00$ & $9.9310$e$+00$ & $3.4680$e$-01$ & $9.6760$ & $0.000$e$+00$ & -3 & main & $0.087$ & $0.383$ & $0.468$ \\
		39 & $0.0000$e$+00$ & $3.3100$e$+01$ & $6.1190$e$+00$ & $1.3270$e$-01$ & $0.0000$ & $1.300$e$+00$ & -1 & main & $0.093$ & $0.525$ & $0.680$ \\
	\enddata
	\tablecomments{ Second set of \WD\ extended data columns. 
	The first column, \code{WDID}, is not actually repeated in the table but is repeated here for clarity. 
	The columns \code{Ell}, \code{Sbc}, \code{Irr}, and \code{AGN} are the template scales, $a_\mathrm{E}$, $a_\mathrm{S}$, $a_\mathrm{I}$, and $a_\mathrm{A}$ in Equation~\ref{eqn:sedform}. 
	The units quoted are the overall normalization given for the templates in \cite{Assef:2010}. 
	\code{AGN\_EBmV} is the excess extinction, $E(B-V)$, applied to the AGN template. 
	\code{chisqr} is the $\chi^2$ of the model from Equation~\ref{eqn:sedchisqr}, and \code{Ndf} is the formal number of degrees of freedom in the model (number of filters minus five). 
	\code{FitMode} describes whether \code{AGN\_EBmV} was allowed to vary (``main") or not (``alt").
	\code{z\_min} is the closest redshift at which the galaxy satisfies upper flux cuts, \code{z\_mid} is the redshift at which it satisfies the middle flux cut, and \code{z\_max} is the farthest redshift at which the galaxy satisfies the lower flux cuts.  }
	\label{tbl:dat:modWD2}
\end{deluxetable}

\floattable
\begin{deluxetable}{rrrrr}
	\tabletypesize{\scriptsize}
	\tablewidth{0.72\textwidth}
	\tablecaption{Excerpt from 6dFGS Extended Data, Column Set 1}
	\tablehead{ \colhead{\code{Ra}} & \colhead{\code{Dec}} & \colhead{\code{6dFGS\_ID}} & 
		\colhead{\code{GALEX\_ID}} & \colhead{\code{AllWISE\_ID}}  \\
		\colhead{$\deg$} & \colhead{$\deg$} &  &
		&   }
	\startdata
		$359.499210$ & $-28.958080$ & 7 & 6380767410811569507 & 0000m288\_ac51-016147 \\
		$359.369120$ & $-29.047580$ & 11 & 6380767411885310513 & 0000m288\_ac51-013523 \\
		$358.042580$ & $-29.079060$ & 19 & 6380767412959052516 & 3582m288\_ac51-016561 \\
		$358.872330$ & $-27.883440$ & 37 & 6380767402221635495 & 3583m273\_ac51-000009 \\
		$359.041330$ & $-27.466580$ & 39 & 6380767391484215299 & 3583m273\_ac51-024746 \\
		$0.059370$ & $-26.730940$ & 52 & 6380767390412573430 & 0000m273\_ac51-059368 \\
	\enddata
	\tablecomments{ First set of 6dFGS extended data columns. 
	The first two columns are the right ascension and declination, in J2000 decimal degrees. 
	The \code{6dFGS\_ID} column is an integer index uniquely assigned to targets in the 6dFGS survey. 
	\code{GALEX\_ID} is a uniquely identifying integer assigned in GALEX gr7 to the source (\code{null} if not matched). 
	\code{AllWISE\_ID} is the \href{http://wise2.ipac.caltech.edu/docs/release/allwise/expsup/sec2\_1a.html\#source\_id}{\code{source\_id}} assigned to the source in the AllWISE survey.  }
	\label{tbl:dat:mod6d1}
\end{deluxetable}

\floattable
\begin{deluxetable}{r|rrrrrrrrrr}
	\tabletypesize{\scriptsize}
	\rotate
	\tablewidth{0.77\textheight}
	\tablecaption{Excerpt from 6dFGS Extended Data, Column Set 2}
	\tablehead{ \colhead{\code{6dFGS\_ID}} & 
		\colhead{\code{Ell}} & \colhead{\code{Sbc}} & \colhead{\code{Irr}} & \colhead{\code{AGN}} & 
		\colhead{\code{AGN\_EBmV}} & \colhead{\code{chisqr}} & \colhead{\code{Ndf}} & 
		\colhead{\code{FitMode}} & 
		\colhead{\code{z\_min}} & \colhead{\code{z\_max}} \\
		 & 
		\colhead{$10^{10}\, L_\odot$} & \colhead{$10^{10}\, L_\odot$} & \colhead{$10^{10}\, L_\odot$} & \colhead{$10^{10}\, L_\odot$} &
		\colhead{$\operatorname{mag}$} &  & 
		& 
		&  & }
	\startdata
		7 & $2.5270$e$+01$ & $0.0000$e$+00$ & $3.5290$e$-01$ & $0.0000$e$+00$ & $0.0000$ & $2.40$e$+02$ & $1$ & alt & $0.053$ & $0.102$ \\
		11 & $2.5500$e$+01$ & $0.0000$e$+00$ & $2.5950$e$-01$ & $1.8700$e$-03$ & $2.0360$ & $1.00$e$+02$ & $2$ & main & $0.056$ & $0.106$ \\
		19 & $8.4310$e$-01$ & $0.0000$e$+00$ & $0.0000$e$+00$ & $0.0000$e$+00$ & $0.0000$ & $5.40$e$+04$ & $3$ & alt & $0.045$ & $0.085$ \\
		37 & $7.0290$e$-01$ & $0.0000$e$+00$ & $0.0000$e$+00$ & $0.0000$e$+00$ & $0.0000$ & $2.70$e$+03$ & $2$ & alt & $0.044$ & $0.085$ \\
		39 & $3.0380$e$-01$ & $0.0000$e$+00$ & $6.4680$e$-02$ & $8.6020$e$-04$ & $0.0000$ & $3.40$e$+01$ & $2$ & main & $0.007$ & $0.014$ \\
		52 & $1.8140$e$+00$ & $0.0000$e$+00$ & $0.0000$e$+00$ & $0.0000$e$+00$ & $0.0000$ & $1.70$e$+04$ & $3$ & alt & $0.042$ & $0.080$ \\
	\enddata
	\tablecomments{ Second set of 6dFGS extended data columns. 
	The first column, \code{6dFGS\_ID}, is not actually repeated in the table but is repeated here for clarity. 
	The columns \code{Ell}, \code{Sbc}, \code{Irr}, and \code{AGN} are the template scales, $a_\mathrm{E}$, $a_\mathrm{S}$, $a_\mathrm{I}$, and $a_\mathrm{A}$ in Equation~\ref{eqn:sedform}. 
	The units quoted are the overall normalization given for the templates in \cite{Assef:2010}. 
	\code{AGN\_EBmV} is the excess extinction, $E(B-V)$, applied to the AGN template. 
	\code{chisqr} is the $\chi^2$ of the model from Equation~\ref{eqn:sedchisqr}, and \code{Ndf} is the formal number of degrees of freedom in the model (number of filters minus five). 
	\code{FitMode} describes whether \code{AGN\_EBmV} was allowed to vary (``main") or not (``alt").
	\code{z\_min} is the closest redshift at which the galaxy satisfies upper flux cuts, and \code{z\_max} is the farthest redshift at which the galaxy satisfies the lower flux cuts. }
	\label{tbl:dat:mod6d2}
\end{deluxetable}

\floattable
\begin{deluxetable}{rrrr}
	\tabletypesize{\scriptsize}
	\tablewidth{0.72\textwidth}
	\tablecaption{Excerpt from SDSS Extended Data, Column Set 1}
	\tablehead{ \colhead{\code{Ra}} & \colhead{\code{Dec}} & \colhead{\code{SDSS\_ID}} & 
		\colhead{\code{AllWISE\_ID}}  \\
		\colhead{$\deg$} & \colhead{$\deg$} &  &
		  }
	\startdata
		$54.936790$ & $0.216800$ & 468504134002173952 & 0544p000\_ac51-047826 \\
		$57.025340$ & $0.208850$ & 1398488404370417664 & 0574p000\_ac51-036206 \\
		$57.296590$ & $0.185310$ & 1398496375829719040 & 0574p000\_ac51-036226 \\
		$57.442290$ & $0.158840$ & 1398494451684370432 & 0574p000\_ac51-038743 \\
		$57.452670$ & $0.044340$ & 1398499399486695424 & 0574p000\_ac51-027223 \\
		$57.490400$ & $0.074350$ & 1398501598509950976 & 0574p000\_ac51-027204 \\
	\enddata
	\tablecomments{ First set of SDSS extended data columns. 
	The first two columns are the right ascension and declination, in J2000 decimal degrees. 
	The \code{SDSS\_ID} column is an integer index uniquely assigned to targets in the SDSS survey (comes from \code{specObjID} column of the \code{SpecObj} table in the SDSS data release 10 context of CasJobs). 
	\code{AllWISE\_ID} is the \href{http://wise2.ipac.caltech.edu/docs/release/allwise/expsup/sec2\_1a.html\#source\_id}{\code{source\_id}} assigned to the source in the AllWISE survey.  }
	\label{tbl:dat:modSD1}
\end{deluxetable}

\floattable
\begin{deluxetable}{r|rrrrrrrrrr}
	\tabletypesize{\scriptsize}
	\rotate
	\tablewidth{0.85\textheight}
	\tablecaption{Excerpt from SDSS Extended Data, Column Set 2}
	\tablehead{ \colhead{\code{SDSS\_ID}} & 
		\colhead{\code{Ell}} & \colhead{\code{Sbc}} & \colhead{\code{Irr}} & \colhead{\code{AGN}} & 
		\colhead{\code{AGN\_EBmV}} & \colhead{\code{chisqr}} & \colhead{\code{Ndf}} & 
		\colhead{\code{FitMode}} &
		\colhead{\code{z\_min}} & \colhead{\code{z\_max}} \\
		 & 
		\colhead{$10^{10}\, L_\odot$} & \colhead{$10^{10}\, L_\odot$} & \colhead{$10^{10}\, L_\odot$} & \colhead{$10^{10}\, L_\odot$} &
		\colhead{$\operatorname{mag}$} &  &  &
		&
		& 
		}
	\startdata
		468504134002173952 & $3.1090$e$+01$ & $0.0000$e$+00$ & $9.6020$e$-01$ & $0.0000$e$+00$ & $0.0000$ & $5.700$e$+02$ & $5$ & main & $0.001$ & $0.232$ \\
		1398488404370417664 & $1.2430$e$+00$ & $0.0000$e$+00$ & $5.3510$e$-01$ & $0.0000$e$+00$ & $0.0000$ & $1.400$e$+03$ & $7$ & alt & $0.001$ & $0.085$ \\
		1398496375829719040 & $3.2860$e$+01$ & $0.0000$e$+00$ & $4.5820$e$+00$ & $0.0000$e$+00$ & $0.0000$ & $1.400$e$+03$ & $7$ & alt & $0.001$ & $0.244$ \\
		1398494451684370432 & $4.4810$e$+01$ & $0.0000$e$+00$ & $2.9780$e$+00$ & $0.0000$e$+00$ & $0.0000$ & $1.400$e$+03$ & $5$ & alt & $0.001$ & $0.291$ \\
		1398499399486695424 & $7.1240$e$+00$ & $1.0600$e$+01$ & $4.5760$e$-01$ & $3.1490$e$-01$ & $13.0300$ & $2.000$e$+02$ & $7$ & main & $0.001$ & $0.154$ \\
		1398501598509950976 & $3.2950$e$+01$ & $0.0000$e$+00$ & $1.0460$e$+00$ & $0.0000$e$+00$ & $0.0000$ & $3.600$e$+02$ & $5$ & main & $0.001$ & $0.239$ \\
	\enddata
	\tablecomments{ Second set of SDSS extended data columns. 
	The first column, \code{SDSS\_ID}, is not actually repeated in the table but is repeated here for clarity. 
	The columns \code{Ell}, \code{Sbc}, \code{Irr}, and \code{AGN} are the template scales, $a_\mathrm{E}$, $a_\mathrm{S}$, $a_\mathrm{I}$, and $a_\mathrm{A}$ in Equation~\ref{eqn:sedform}. 
	The units quoted are the overall normalization given for the templates in \cite{Assef:2010}. 
	\code{AGN\_EBmV} is the excess extinction, $E(B-V)$, applied to the AGN template. 
	\code{chisqr} is the $\chi^2$ of the model from Equation~\ref{eqn:sedchisqr}, and \code{Ndf} is the formal number of degrees of freedom in the model (number of filters minus five). 
	\code{FitMode} describes whether \code{AGN\_EBmV} was allowed to vary (``main") or not (``alt").
	\code{z\_min} is the closest redshift at which the galaxy satisfies upper flux cuts, and \code{z\_max} is the farthest redshift at which the galaxy satisfies the lower flux cuts. }
	\label{tbl:dat:modSD2}
\end{deluxetable}

\floattable
\begin{deluxetable}{rrrr}
	\tabletypesize{\scriptsize}
	\tablewidth{0.48\textwidth}
	\tablecaption{Excerpt from GAMA Extended Data, Column Set 1}
	\tablehead{ \colhead{\code{Ra}} & \colhead{\code{Dec}} & \colhead{\code{GAMA\_ID}} & 
		\colhead{\code{AllWISE\_ID}}  \\
		\colhead{$\deg$} & \colhead{$\deg$} &
		&  }
	\startdata
		$174.022810$ & $0.705940$ & 6806 & 1739p000\_ac51-051576 \\
		$174.100730$ & $0.658910$ & 6808 & 1739p000\_ac51-051657 \\
		$174.184930$ & $0.709040$ & 6826 & 1739p000\_ac51-049388 \\
		$174.302790$ & $0.789990$ & 6837 & 1739p015\_ac51-002444 \\
		$174.346900$ & $0.696450$ & 6840 & 1739p000\_ac51-049335 \\
		$174.396030$ & $0.820770$ & 6844 & 1739p015\_ac51-002401 \\
	\enddata
	\tablecomments{ First set of GAMA extended data columns. 
	The first two columns are the right ascension and declination, in J2000 decimal degrees. 
	The \code{GAMA\_ID} column is an integer index uniquely assigned to targets in the GAMA survey. 
	\code{AllWISE\_ID} is the \href{http://wise2.ipac.caltech.edu/docs/release/allwise/expsup/sec2\_1a.html\#source\_id}{\code{source\_id}} assigned to the source in the AllWISE survey. }
	\label{tbl:dat:modGA1}
\end{deluxetable}

\floattable
\begin{deluxetable}{r|rrrrrrrrrr}
	\tabletypesize{\scriptsize}
	\rotate
	\tablewidth{0.76\textheight}
	\tablecaption{Excerpt from GAMA Extended Data, Column Set 2}
	\tablehead{ \colhead{\code{GAMA\_ID}} & 
		\colhead{\code{Ell}} & \colhead{\code{Sbc}} & \colhead{\code{Irr}} & \colhead{\code{AGN}} & 
		\colhead{\code{AGN\_EBmV}} & \colhead{\code{chisqr}} & \colhead{\code{Ndf}} & 
		\colhead{\code{FitMode}} & 
		\colhead{\code{z\_min}} & \colhead{\code{z\_max}} \\
		 & 
		\colhead{$10^{10}\, L_\odot$} & \colhead{$10^{10}\, L_\odot$} & \colhead{$10^{10}\, L_\odot$} & \colhead{$10^{10}\, L_\odot$} &
		\colhead{$\operatorname{mag}$} &  & 
		& 
		&  & }
	\startdata
		6806 & $1.1150$e$+01$ & $2.7000$e$+01$ & $2.7750$e$+00$ & $4.8930$e$-02$ & $0.1746$ & $1.900$e$+02$ & $8$ & main & $0.052$ & $0.383$ \\
		6808 & $1.0400$e$+01$ & $0.0000$e$+00$ & $5.2290$e$-01$ & $3.2660$e$-03$ & $0.0000$ & $2.300$e$+02$ & $6$ & main & $0.031$ & $0.246$ \\
		6826 & $2.7550$e$+00$ & $0.0000$e$+00$ & $3.2140$e$-01$ & $0.0000$e$+00$ & $0.0000$ & $3.600$e$+03$ & $7$ & alt & $0.015$ & $0.130$ \\
		6837 & $1.2690$e$+00$ & $0.0000$e$+00$ & $3.7570$e$-01$ & $0.0000$e$+00$ & $0.0000$ & $1.100$e$+03$ & $7$ & alt & $0.014$ & $0.122$ \\
		6840 & $1.1440$e$+01$ & $0.0000$e$+00$ & $3.8100$e$-01$ & $3.0570$e$-02$ & $0.2704$ & $5.300$e$+02$ & $7$ & main & $0.032$ & $0.248$ \\
		6844 & $6.3390$e$+00$ & $0.0000$e$+00$ & $1.1100$e$-01$ & $0.0000$e$+00$ & $0.0000$ & $1.100$e$+03$ & $7$ & alt & $0.024$ & $0.192$ \\
	\enddata
	\tablecomments{ Second set of GAMA extended data columns. 
	The first column, \code{GAMA\_ID}, is not actually repeated in the table but is repeated here for clarity. 
	The columns \code{Ell}, \code{Sbc}, \code{Irr}, and \code{AGN} are the template scales, $a_\mathrm{E}$, $a_\mathrm{S}$, $a_\mathrm{I}$, and $a_\mathrm{A}$ in Equation~\ref{eqn:sedform}. 
	The units quoted are the overall normalization given for the templates in \cite{Assef:2010}. 
	\code{AGN\_EBmV} is the excess extinction, $E(B-V)$, applied to the AGN template. 
	\code{chisqr} is the $\chi^2$ of the model from Equation~\ref{eqn:sedchisqr}, and \code{Ndf} is the formal number of degrees of freedom in the model (number of filters minus five). 
	\code{FitMode} describes whether \code{AGN\_EBmV} was allowed to vary (``main") or not (``alt").
	\code{z\_min} is the closest redshift at which the galaxy satisfies upper flux cuts, and \code{z\_max} is the farthest redshift at which the galaxy satisfies the lower flux cuts. }
	\label{tbl:dat:modGA2}
\end{deluxetable}

\floattable
\begin{deluxetable}{rrrrrr}
	\tabletypesize{\scriptsize}
	\tablewidth{0.8\textwidth}
	\tablecaption{Excerpt from AGES Extended Data, Column Set 1}
	\tablehead{ \colhead{\code{Ra}} & \colhead{\code{Dec}} & \colhead{\code{AGES\_ROW}} & 
		\colhead{\code{NDWFS\_ID}} & \colhead{\code{SDSS\_ID}} & \colhead{\code{AllWISE\_ID}}  \\
		\colhead{$\deg$} & \colhead{$\deg$} &  &
		& &  }
	\startdata
		$216.393850$ & $32.806960$ & 6 & 346042 & 1237662684146041043 & 2159p333\_ac51-002610 \\
		$216.548230$ & $32.807660$ & 11 & 346533 & 1237662684146106652 & 2159p333\_ac51-002915 \\
		$216.818990$ & $32.809900$ & 26 & 348131 & null & 2159p333\_ac51-000046 \\
		$216.245250$ & $32.812800$ & 53 & 350281 & 1237664852570800769 & 2159p333\_ac51-002806 \\
		$217.374410$ & $32.814140$ & 64 & 351325 & 1237664853108064555 & 2177p333\_ac51-008367 \\
		$216.179020$ & $32.814450$ & 68 & 351568 & 1237664852570800403 & 2159p333\_ac51-005461 \\
	\enddata
	\tablecomments{ First set of AGES extended data columns. 
	The first two columns are the right ascension and declination, in J2000 decimal degrees. 
	AGES does not have an identifier for its sources, but the plain text tables in \cite{AGES} have corresponding rows. 
	\code{AGES\_ROW} column contains the identity of the row the galaxy was published in, starting from 0. 
	\code{NDWFS\_ID} is a uniquely identifying integer assigned in NDWFS to the source (\code{null} if not matched). 
	\code{SDSS\_ID} is a uniquely identifying integer assigned in SDSS to the matching source (\code{null} if not matched). 
	\code{AllWISE\_ID} is the \href{http://wise2.ipac.caltech.edu/docs/release/allwise/expsup/sec2\_1a.html\#source\_id}{\code{source\_id}} assigned to the source in the AllWISE survey.  }
	\label{tbl:dat:modAG1}
\end{deluxetable}

\floattable
\begin{deluxetable}{r|rrrrrrrrrrr}
	\tabletypesize{\scriptsize}
	\rotate
	\tablewidth{0.83\textheight}
	\tablecaption{Excerpt from AGES Extended Data, Column Set 2}
	\tablehead{ \colhead{\code{AGES\_ROW}} & 
		\colhead{\code{Ell}} & \colhead{\code{Sbc}} & \colhead{\code{Irr}} & \colhead{\code{AGN}} & 
		\colhead{\code{AGN\_EBmV}} & \colhead{\code{chisqr}} & \colhead{\code{Ndf}} & 
		\colhead{\code{FitMode}} &
		\colhead{\code{z\_min}} & \colhead{\code{z\_mid}} & \colhead{\code{z\_max}} \\
		 & 
		\colhead{$10^{10}\, L_\odot$} & \colhead{$10^{10}\, L_\odot$} & \colhead{$10^{10}\, L_\odot$} & \colhead{$10^{10}\, L_\odot$} &
		\colhead{$\operatorname{mag}$} &  &  & 
		& 
		&  &  }
	\startdata
		6 & $1.6020$e$+00$ & $5.7740$e$+00$ & $0.0000$e$+00$ & $0.0000$e$+00$ & $0.0000$ & $4.70$e$+03$ & $4$ & alt & $0.006$ & $0.200$ & $0.243$ \\
		11 & $3.7850$e$+00$ & $0.0000$e$+00$ & $0.0000$e$+00$ & $9.4590$e$-02$ & $0.2121$ & $2.00$e$+02$ & $4$ & main & $0.006$ & $0.190$ & $0.276$ \\
		26 & $7.4010$e$+01$ & $0.0000$e$+00$ & $0.0000$e$+00$ & $0.0000$e$+00$ & $0.0000$ & $4.90$e$+03$ & $4$ & alt & $0.026$ & $0.580$ & $0.865$ \\
		53 & $2.5880$e$+01$ & $0.0000$e$+00$ & $2.8970$e$+00$ & $0.0000$e$+00$ & $0.0000$ & $4.40$e$+01$ & $4$ & main & $0.017$ & $0.453$ & $0.745$ \\
		64 & $1.0010$e$+01$ & $2.7350$e$+00$ & $0.0000$e$+00$ & $9.0590$e$-02$ & $0.1298$ & $2.90$e$+02$ & $4$ & main & $0.010$ & $0.293$ & $0.475$ \\
		68 & $1.5140$e$+01$ & $1.1810$e$+00$ & $1.1090$e$-01$ & $0.0000$e$+00$ & $0.0000$ & $4.20$e$+02$ & $1$ & alt & $0.012$ & $0.333$ & $0.481$ \\
	\enddata
	\tablecomments{ Second set of AGES extended data columns. 
	The first column, \code{AGES\_ROW}, is not actually repeated in the table but is repeated here for clarity. 
	The columns \code{Ell}, \code{Sbc}, \code{Irr}, and \code{AGN} are the template scales, $a_\mathrm{E}$, $a_\mathrm{S}$, $a_\mathrm{I}$, and $a_\mathrm{A}$ in Equation~\ref{eqn:sedform}. 
	The units quoted are the overall normalization given for the templates in \cite{Assef:2010}. 
	\code{AGN\_EBmV} is the excess extinction, $E(B-V)$, applied to the AGN template. 
	\code{chisqr} is the $\chi^2$ of the model from Equation~\ref{eqn:sedchisqr}, and \code{Ndf} is the formal number of degrees of freedom in the model (number of filters minus five). 
	\code{FitMode} describes whether \code{AGN\_EBmV} was allowed to vary (``main") or not (``alt").
	\code{z\_min} is the closest redshift at which the galaxy satisfies upper flux cuts, \code{z\_mid} is the redshift at which it satisfies the middle flux cut, and \code{z\_max} is the farthest redshift at which the galaxy satisfies the lower flux cuts. }
	\label{tbl:dat:modAG2}
\end{deluxetable}

\floattable
\begin{deluxetable}{rrrrrrr}
	\tabletypesize{\scriptsize}
	\tablewidth{0.86\textwidth}
	\tablecaption{Excerpt from zCOSMOS Extended Data, Column Set 1}
	\tablehead{ \colhead{\code{Ra}} & \colhead{\code{Dec}} & \colhead{\code{zCOS\_ID}} & 
		\colhead{\code{COS\_ID}} & \colhead{\code{SCOS\_ID}} & \colhead{\code{SDSS\_ID}} & \colhead{\code{AllWISE\_ID}}  \\
		\colhead{$\deg$} & \colhead{$\deg$} & &
		&  &  }
	\startdata
		$150.502790$ & $1.877650$ & 700137 & 507130 & null & null & 1497p015\_ac51-048699 \\
		$150.280590$ & $2.021280$ & 700529 & null & null & 1237653664722125224 & 1497p015\_ac51-051213 \\
		$150.122600$ & $2.108540$ & 700585 & 768236 & null & null & 1497p015\_ac51-054126 \\
		$150.183040$ & $2.028990$ & 700587 & null & 128673 & null & 1497p015\_ac51-053799 \\
		$150.393000$ & $2.342770$ & 701269 & 1213568 & 199260 & null & 1497p030\_ac51-000271 \\
		$150.653290$ & $1.625360$ & 800270 & 67120 & 41381 & 1237653664185385269 & 1512p015\_ac51-036648 \\
	\enddata
	\tablecomments{ First set of zCOSMOS extended data columns. 
	The first two columns are the right ascension and declination, in J2000 decimal degrees. 
	The \code{zCOS\_ID} column is an integer index uniquely assigned to targets in the zCOSMOS survey. 
	\code{COS\_ID} is a uniquely identifying integer assigned in \cite{Capak:2007} to the matching source (\code{null} if not matched). 
	\code{SCOS\_ID} is a uniquely identifying integer assigned in SCOSMOS to the matching source (\code{null} if not matched). 
	\code{AllWISE\_ID} is the \href{http://wise2.ipac.caltech.edu/docs/release/allwise/expsup/sec2\_1a.html\#source\_id}{\code{source\_id}} assigned to the source in the AllWISE survey. }
	\label{tbl:dat:modzC1}
\end{deluxetable}

\floattable
\begin{deluxetable}{r|rrrrrrrrrr}
	\tabletypesize{\scriptsize}
	\rotate
	\tablewidth{0.8\textheight}
	\tablecaption{Excerpt from zCOSMOS Extended Data, Column Set 2}
	\tablehead{ \colhead{\code{zCOS\_ID}} & 
		\colhead{\code{Ell}} & \colhead{\code{Sbc}} & \colhead{\code{Irr}} & \colhead{\code{AGN}} & 
		\colhead{\code{AGN\_EBmV}} & \colhead{\code{chisqr}} & \colhead{\code{Ndf}} & 
		\colhead{\code{FitMode}} &
		\colhead{\code{z\_min}} & \colhead{\code{z\_max}} \\
		 & 
		\colhead{$10^{10}\, L_\odot$} & \colhead{$10^{10}\, L_\odot$} & \colhead{$10^{10}\, L_\odot$} & \colhead{$10^{10}\, L_\odot$} &
		\colhead{$\operatorname{mag}$} &  &  &
		& 
		&  }
	\startdata
		700137 & $1.1050$e$+01$ & $6.2610$e$+00$ & $0.0000$e$+00$ & $3.6830$e$+00$ & $1.0500$ & $4.900$e$+02$ & $8$ & main & $0.069$ & $0.885$ \\
		700529 & $0.0000$e$+00$ & $8.1030$e$+00$ & $0.0000$e$+00$ & $4.4850$e$-01$ & $13.0300$ & $1.600$e$-05$ & $-2$ & main & $0.016$ & $0.283$ \\
		700585 & $5.2150$e$+00$ & $0.0000$e$+00$ & $0.0000$e$+00$ & $3.2440$e$+00$ & $0.6921$ & $1.400$e$+03$ & $11$ & main & $0.043$ & $0.733$ \\
		700587 & $1.2620$e$+01$ & $0.0000$e$+00$ & $1.6640$e$+00$ & $0.0000$e$+00$ & $0.0000$ & $1.600$e$+03$ & $2$ & alt & $0.021$ & $0.471$ \\
		701269 & $0.0000$e$+00$ & $2.8730$e$+01$ & $7.7650$e$-01$ & $0.0000$e$+00$ & $0.0000$ & $9.300$e$+03$ & $17$ & alt & $0.088$ & $0.574$ \\
		800270 & $0.0000$e$+00$ & $2.8730$e$+01$ & $7.7650$e$-01$ & $0.0000$e$+00$ & $0.0000$ & $9.300$e$+03$ & $17$ & alt & $0.088$ & $0.574$ \\
	\enddata
	\tablecomments{ Second set of zCOSMOS extended data columns. 
	The first column, \code{zCOS\_ID}, is not actually repeated in the table but is repeated here for clarity. 
	The columns \code{Ell}, \code{Sbc}, \code{Irr}, and \code{AGN} are the template scales, $a_\mathrm{E}$, $a_\mathrm{S}$, $a_\mathrm{I}$, and $a_\mathrm{A}$ in Equation~\ref{eqn:sedform}. 
	The units quoted are the overall normalization given for the templates in \cite{Assef:2010}. 
	\code{AGN\_EBmV} is the excess extinction, $E(B-V)$, applied to the AGN template. 
	\code{chisqr} is the $\chi^2$ of the model from Equation~\ref{eqn:sedchisqr}, and \code{Ndf} is the formal number of degrees of freedom in the model (number of filters minus five). 
	\code{FitMode} describes whether \code{AGN\_EBmV} was allowed to vary (``main") or not (``alt").
	\code{z\_min} is the closest redshift at which the galaxy satisfies upper flux cuts, and \code{z\_max} is the farthest redshift at which the galaxy satisfies the lower flux cuts. }
	\label{tbl:dat:modzC2}
\end{deluxetable}

\section{Discussion}\label{sec:disc}
The data gathered and characterized here was collected primarily to use in measuring the $2.4\micron$ luminosity function of all galaxies back to a redshift of $z = 1$, as is done in this work's companion paper LW17III. 
The main purpose of this work is to describe, in detail, the cuts made to the data and the characteristics of the resulting set. 
This process is an essential component in evaluating the sensitivity of the measurements carried out in LW17III and in making the data presented here both auditable and extendable.  
The multi-wavelength data sets available for most of the surveys covered here are extensive, and a more sophisticated spectro-luminosity functional analysis than what is in LW17III should be possible if a fast and deterministic high dimension Gaussian integrator can be developed.

\acknowledgements

We would like to thank the \textbf{\WISE} team.\\
This publication makes use of data products from the Wide-field Infrared Survey Explorer, which is a joint project of the University of California, Los Angeles, and the Jet Propulsion Laboratory/California Institute of Technology, and NEOWISE, which is a project of the Jet Propulsion Laboratory/California Institute of Technology. WISE and NEOWISE are funded by the National Aeronautics and Space Administration.

We would like to thank the \textbf{SDSS} team.\\
Funding for SDSS-III has been provided by the Alfred P. Sloan Foundation, the Participating Institutions, the National Science Foundation, and the U.S. Department of Energy Office of Science. The SDSS-III web site is http://www.sdss3.org/.

SDSS-III is managed by the Astrophysical Research Consortium for the Participating Institutions of the SDSS-III Collaboration including the University of Arizona, the Brazilian Participation Group, Brookhaven National Laboratory, Carnegie Mellon University, University of Florida, the French Participation Group, the German Participation Group, Harvard University, the Instituto de Astrofisica de Canarias, the Michigan State/Notre Dame/JINA Participation Group, Johns Hopkins University, Lawrence Berkeley National Laboratory, Max Planck Institute for Astrophysics, Max Planck Institute for Extraterrestrial Physics, New Mexico State University, New York University, Ohio State University, Pennsylvania State University, University of Portsmouth, Princeton University, the Spanish Participation Group, University of Tokyo, University of Utah, Vanderbilt University, University of Virginia, University of Washington, and Yale University.

We would like to thank the \textbf{GAMA} team.\\
GAMA is a joint European-Australasian project based around a spectroscopic campaign using the Anglo-Australian Telescope. The GAMA input catalogue is based on data taken from the Sloan Digital Sky Survey and the UKIRT Infrared Deep Sky Survey. Complementary imaging of the GAMA regions is being obtained by a number of independent survey programs including GALEX MIS, VST KiDS, VISTA VIKING, WISE, Herschel-ATLAS, GMRT and ASKAP providing UV to radio coverage. GAMA is funded by the STFC (UK), the ARC (Australia), the AAO, and the participating institutions. The GAMA website is http://www.gama-survey.org/ .

Based on observations made with ESO Telescopes at the La Silla or Paranal Observatories under programme ID 175.A-0839.

We would like to thank the \textbf{2MASS} team.\\
This publication makes use of data products from the Two Micron All Sky Survey, which is a joint project of the University of Massachusetts and the Infrared Processing and Analysis Center/California Institute of Technology, funded by the National Aeronautics and Space Administration and the National Science Foundation.

We would like to thank the \textbf{MAST} team.\\
Some/all of the data presented in this paper were obtained from the Mikulski Archive for Space Telescopes (MAST). STScI is operated by the Association of Universities for Research in Astronomy, Inc., under NASA contract NAS5-26555. Support for MAST for non-HST data is provided by the NASA Office of Space Science via grant NNX13AC07G and by other grants and contracts.

We would like to thank the \textbf{NDWFS} team.\\
This work made use of images and/or data products provided by the NOAO Deep Wide-Field Survey (Jannuzi and Dey 1999; Jannuzi et al. 2005; Dey et al. 2005), which is supported by the National Optical Astronomy Observatory (NOAO). NOAO is operated by AURA, Inc., under a cooperative agreement with the National Science Foundation.

We would like to thank the \textbf{IPAC} team.\\
This research has made use of the NASA/ IPAC Infrared Science Archive, which is operated by the Jet Propulsion Laboratory, California Institute of Technology, under contract with the National Aeronautics and Space Administration.

We would like to thank the \textbf{GALEX} team.\\
Based on observations made with the NASA Galaxy Evolution Explorer. 
GALEX is operated for NASA by the California Institute of Technology under NASA contract NAS5-98034.

We would also like to thank the teams behind \textbf{6dFGS}, \textbf{AGES}, \textbf{zCOSMOS}, \textbf{SDWFS}, and \textbf{COSMOS}.

RJA was supported by FONDECYT grant number 1151408.

\bibliography{DataBib}

\end{document}